\documentclass[pra,aps,twocolumn,superscriptaddress,showpacs,eqsecnum,nofootinbib]{revtex4-2}

\usepackage{amsmath}
\usepackage{amssymb}
\usepackage{graphicx}
\usepackage{color}
\usepackage{hyperref}
\usepackage{ulem}

\begin{document}

\title{Coupled dynamics of spin qubits in optical dipole microtraps}

\author{L.V. Gerasimov}
\affiliation{Quantum Technology Centre, Faculty of Physics, M.V. Lomonosov Moscow State University, Leninskiye Gory 1-35, 119991, Moscow, Russia}
\affiliation{Center for Advanced Studies, Peter the Great St. Petersburg Polytechnic University, 195251, St. Petersburg, Russia}
\author{R.R. Yusupov}
\affiliation{Quantum Technology Centre, Faculty of Physics, M.V. Lomonosov Moscow State University, Leninskiye Gory 1-35, 119991, Moscow, Russia}
\author{A.D. Moiseevsky}
\affiliation{Quantum Technology Centre, Faculty of Physics, M.V. Lomonosov Moscow State University, Leninskiye Gory 1-35, 119991, Moscow, Russia}
\author{I. Vybornyi}
\affiliation{Institut f\"ur Theoretische Physik Leibniz Universit\"at Hannover, Appelstra\ss e 2, 30167 Hannover, Germany}
\author{K.S.Tikhonov}
\affiliation{St. Petersburg State University, 199034, St. Petersburg, Russia}
\affiliation{Russian Quantum Center, Skolkovo, Moscow 143025, Russia}
\author{S.P. Kulik}
\affiliation{Quantum Technology Centre, Faculty of Physics, M.V. Lomonosov Moscow State University, Leninskiye Gory 1-35, 119991, Moscow, Russia}
\author{S.S. Straupe}
\affiliation{Quantum Technology Centre, Faculty of Physics, M.V. Lomonosov Moscow State University, Leninskiye Gory 1-35, 119991, Moscow, Russia}
\affiliation{Russian Quantum Center, Skolkovo, Moscow 143025, Russia}
\author{C.I. Sukenik}
\affiliation{Department of Physics, Old Dominion University 4600 Elkhorn Ave. Norfolk, VA 23529 USA}
\author{D.V. Kupriyanov}
\affiliation{Quantum Technology Centre, Faculty of Physics, M.V. Lomonosov Moscow State University, Leninskiye Gory 1-35, 119991, Moscow, Russia}
\affiliation{Department of Physics, Old Dominion University 4600 Elkhorn Ave. Norfolk, VA 23529 USA}

\begin{abstract}
\noindent
Single atoms in dipole microtraps or optical tweezers have recently become a promising platform for quantum computing and simulation. Here we report a detailed theoretical analysis of the physics underlying an implementation of a Rydberg two-qubit gate in such a system -- a cornerstone protocol in quantum computing with single atoms. We focus on a blockade-type entangling gate and consider various decoherence processes limiting its performance in a real system. We provide numerical estimates for the limits on fidelity of the maximally entangled states and predict the full process matrix corresponding to the noisy two-qubit gate. Our methods and results may find implementation in numerical models for simulation and optimization of neutral atom based quantum processors.

\end{abstract}


\maketitle

\section{Introduction}\label{Section_I}


\noindent During the last few decades, the physics of mesoscopic cold and ultracold atomic systems has been continuously progressing and supplying novel ideas for implementations of innovative state-of-the-art quantum technologies \cite{Bloch_Science2017,Schafer_NatureRevPhys2020}. One of the research directions showing extremely impressive progress is quantum computing and simulation with single trapped atoms \cite{Saffman_JPB2016,Browaeys_NaturePhys2020,Whitlock_AVS2021}. Being a paradigmatic quantum system, single atoms provide a convenient physical realization of qubits, which is attractive in many ways -- overall neutrality allows one to controlably switch the interaction on and off, and optical trapping provides a means for assembling large spatially structured atomic arrays with reasonable prospects for further scaling. 

The alkali metals having a single valence electron and convenient combination of optical and microwave spectra can be trapped by tightly focused far-off-resonance light beams (``optical tweezers'') and spaced conveniently for individual addressing. With the technique of holographic beam shaping developed for such experiments \cite{Browaeys_PRX2014,Ahn_NatureComm2016}, two- and three-dimensional tweezers arrays may be constructed \cite{Browaeys_Science2016,Browaeys_Nature2018,Birkl_PRL2019,Ahn_PRA2017,Zhan_PRL2022} providing a means to assemble mesoscopic scale atomic structures, which can be periodically ordered in a plane with a separation of a few microns and with lifetimes reaching up to seconds. Similar techniques were recently developed for alkaline earth atoms \cite{Endres_PRX2018,Endres_NaturePhys2020,Thompson_arxiv2019}.

Such atomic lattices consisting of single neutral atoms confined with the microscopic optical dipole traps provide a promising platform for preparation of conveniently controllable and scalable multi-qubit systems \cite{Kaufman_NaturePhys2021}. This was recognized more than a decade ago and the potential options and experimental capabilities were earlier reviewed in \cite{SaffmanWalkerMolmer2010}. Despite impressive experimental progress since that time, the fidelity of experimentally demonstrated entangling operations is still on the order of 95--97\% \cite{Lukin_PRL2019,Saffman_PRL2019} which is still below the thresholds required for fault-tolerant quantum computing. This situation in quantum computing with neutral atoms has motivated us to perform a comprehensive theoretical analysis of the main physical mechanisms leading to violation of the ideal scenario for an elementary CNOT quantum gate for a pair of hyperfine encoded atomic qubits.

In this paper we mainly focus on the physics of the process and consider a standard configuration of two alkali-metal atoms with qubits encoded in the clock transition of their ground state hyperfine structure. The spin entanglement is induced via the simplest protocol of Rydberg blockade as proposed in \cite{Lukin_PRL2000}. By the detailed examination of such an elementary quantum logic unit we are aiming to clarify the main physical constraints in the coupled system of two qubits and then to search for optimal physical conditions towards its potential scaling up to a multi-qubit configuration. Alternative realizations of entangling gates proposed recently \cite{Saffman_PRA2016,Lukin_PRL2019,Liu_PRAppl2020,Ryabtsev_PRA2016} share the same non-idealities and sources of decoherence and errors, so our analysis remains applicable with minor modifications. One of the main features of our approach is a fully-quantum treatment of the atomic motional degrees of freedom, making the analysis applicable for a full range of temperatures including atoms cooled close to the motional ground state \cite{Regal_PRX2012,Lukin_PRL2013,Andersen_PRA2017}. We also rigorously analyze the limits of the entanglement protocol set by decoherence processes associated not only with the radiative decay of the Rydberg state, but also with the processes of incoherent Rayleigh and Raman scattering via intermediate states used in the two-photon excitation scheme typical for most experiments. We demonstrate how the entanglement loss can be reduced by proper choice of excitation geometry, providing convenient selection rules that minimize the negative contributions of incoherent scattering.

The paper is organized as follows. In Section \ref{Section_II} we give an overview of the idealized dynamical description of the Rydberg blockade protocol adjusted for the implementation of a CZ gate. Then in Section \ref{Section_III} we present our approach incorporating open system dynamics and show how the ideal dynamical process of the CZ gate is affected by spontaneous loss associated with different channels of incoherent scattering. In Section \ref{Section_IV} we present the results of numerical simulations for the fidelity and truth table of the CNOT gate utilizing parameters which are realistic for most currently existing experimental setups.

\section{Entanglement of the spin states of two atoms: dynamical description} \label{Section_II}

\begin{figure}[tp]
\includegraphics[width=8.6cm]{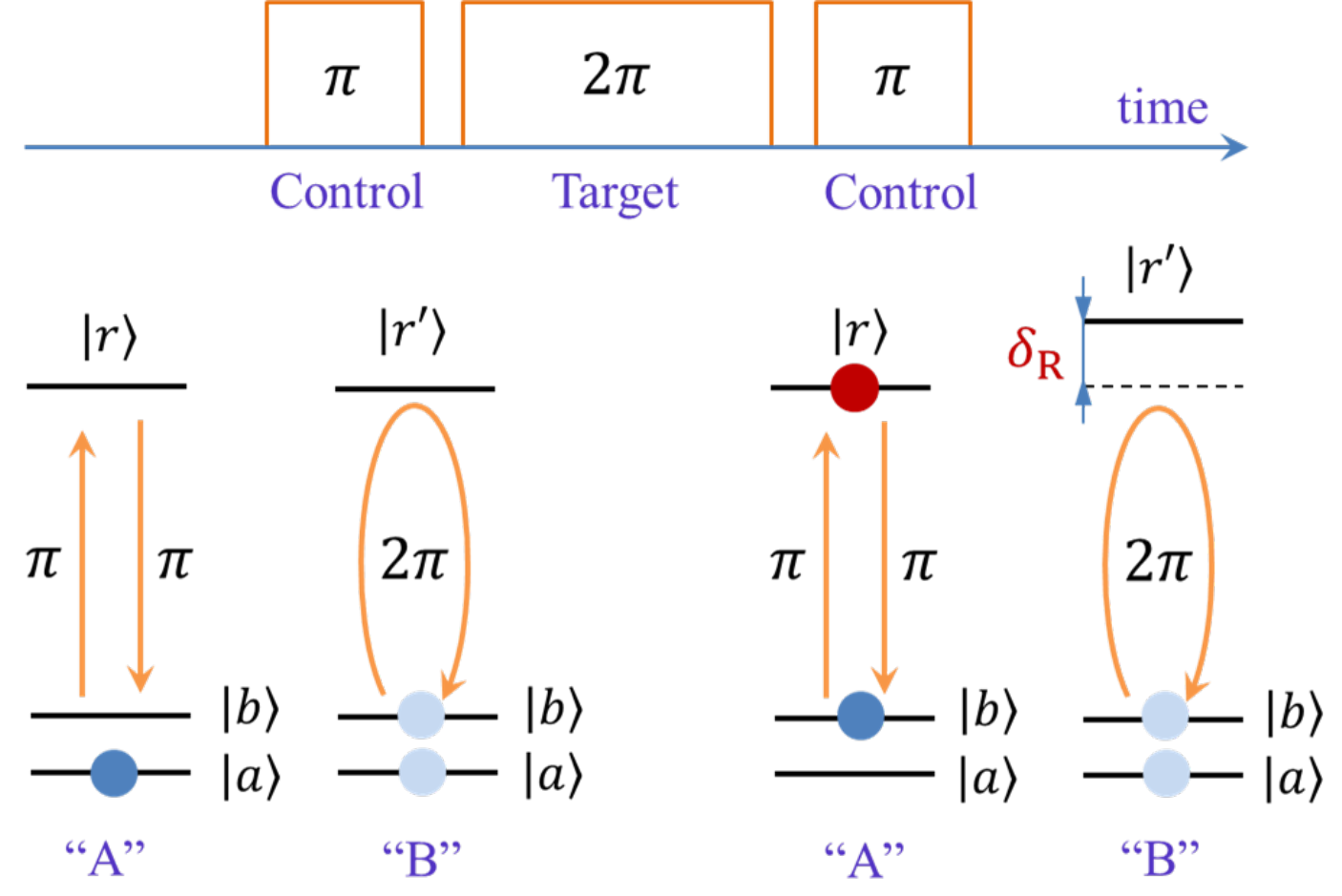}
\caption{The principle of spin entanglement creation via the protocol of Rydberg blockade. If the control atom $A$ occupies a Zeeman state $|a\rangle$ belonging to the lower hyperfine sublevel and the target atom $B$ is in a state $|b\rangle$ belonging to the upper sublevel, the sequence of $\pi-2\pi-\pi$ pulses coupled with the Rydberg states $|r\rangle$ and $|r'\rangle$ changes the phase of the collective spin state by $\pi$. If the atom $A$ occupies the upper spin state $|b\rangle$ its excitation by the $\pi$-pulse to the state $|r\rangle$ shifts the energy level and eliminates the coupling of atom $B$ to the state $|r'\rangle$. The collective spin state again acquires a $\pi$ phase shift. But if both atoms are in the lower spin state $|a\rangle$ the pulse sequence does not change their collective state.}
\label{fig1}%
\end{figure}

\noindent Consider the conventional Rydberg blockade scheme, shown in Fig.~\ref{fig1}, which was proposed in \cite{Lukin_PRL2000}. Let us denote the control atom experiencing the sequence of two two-photon $\pi$-pulses as $A$, and the target atom excited by a $2\pi$-pulse via the transition, which can be blocked by the control atom, as $B$. In an ideal scenario, as a result, the two-particle density matrix $\rho^{AB}\equiv\rho$ of the atomic spin state has to be transformed by a diagonal unitary operator $\mathrm{diag}(+1,-1,-1,-1)$ corresponding to a CZ gate. However, in reality the protocol initiates a set of physical processes disturbing the atomic system and violating the ideal transformation scheme. It is convenient to discuss these processes separately and clarify the theoretical model for each of them independently. 
In this section we address the dynamical part of the protocol treating the system as closed and isolated from the environment, and driven by a specific system Hamiltonian, which we describe below.

\subsection{The system Hamiltonian}\label{Section_IIA}

\noindent In a typical experiment with alkali atoms in a far-off-resonant dipole trap the Rydberg state is slightly anti-trapped. During the protocol of spin entanglement the dipole trap is switched off, the atoms are released in free space, and their motional and internal dynamics are decoupled. Then the atoms are excited by a sequence of short coherent light pulses. 

The system Hamiltonian describing the joint dynamics of the atoms consists of the following contributions:
\begin{equation}
\hat{H}=\hat{H}_0+\sum_{r,r'}\hbar\delta_{R}|r,r'\rangle\langle r,r'|_{AB}+\hat{V}_{\mathrm{eff}}%
\label{2.1}%
\end{equation}
where the undisturbed dynamics is driven by the Hamiltonian $\hat{H}_0$ given by
\begin{equation}
\hat{H}_0=\frac{\hat{\mathbf{p}}_A^2}{2m}+\frac{\hat{\mathbf{p}}_B^2}{2m}+\hat{H}_{A}+\hat{H}_{B}%
\label{2.2}%
\end{equation}
with $\hat{\mathbf{p}}_A$ and $\hat{\mathbf{p}}_B$ being the operators of linear momenta and $\hat{H}_{A}$ and $\hat{H}_{B}$ being the internal Hamiltonians of atoms $A$ and $B$, respectively. Both atoms are physically indistinguishable and have the same mass $m$. 

The critical requirement for the considered system is that being excited in the high-energy Rydberg states $|r\rangle$ the closely spaced atoms $A$ and $B$ separated by a distance of a few microns cannot be considered as independent objects and have a signature of a molecular system. Thus the second term in (\ref{2.1}) corrects the undisturbed Hamiltonian and adds a specific offset $\hbar\delta_R$ to the energy of the doubly excited Rydberg state $|r,r'\rangle_{AB}$, which approximates the behavior of a quasi-molecular orbital at long distances. Here we point out that there is an option that the excited states of the separated atoms can be different by marking one of them with a prime.

Such a model description of the Rydberg blockade can be justified by the following physical arguments. During the entire protocol the atoms are separated in space by dipole traps at a sufficiently large distance of many atomic units, such that the shift $\delta_R$ treated as a far asymptote for the adiabatic potential of a quasi-molecule is insensitive to its slight spatial variations. The atoms are released from the traps to activate the protocol of spin entanglement for a very short time and during this time their locations are not altered significantly. Under these conditions the projection onto the highly excited eigenstates of the system Hamiltonian could be approximated by the product of atomic states having a fixed extra energy shift $\delta_R$ for the double excitation, see \cite{Saffman2008}.

The two-photon excitation process is initiated by two counter-propagating laser beams to minimize the recoil effect in the linear momentum transfer from light to the atoms. The carrier frequency of the first beam $\omega_1$ is quasi-resonant to the manifold of the hyperfine energy structure of the $D_1$-line and the second beam with frequency $\omega_2$ provides the two-photon resonance with the undisturbed Rydberg state $|r\rangle$. \footnote{Note that the hyperfine structure in the Rydberg states is unresolved within the considered microsecond time scale and the spin subsystem can be equivalently described in either the spin decoupled or coupled bases.} With these assumptions we can adiabatically eliminate the dynamics of intermediate states and reduce the two-photon interaction to the effective interaction Hamiltonian

\begin{equation}
 \hat{V}_{\mathrm{eff}}=\hat{V}_{A}(\mathbf{r}_A,t) + \hat{V}_{B}(\mathbf{r}_B,t).%
 \label{2.3}%
\end{equation}
Both contributions are functionally identical and with enumerating the atoms by $X=A,B$ we can specify this part of the Hamiltonian in the position representation for each atom located at point $\mathbf{r}_X$ as
\begin{equation}
\hat{V}_{X}(\mathbf{r}_X,t)=-\frac{\hbar}{2}\,\Omega(\mathbf{r}_X,t)\,%
\mathrm{e}^{-i\omega t+i\mathbf{q}\cdot\mathbf{r}_X}\,|r\rangle\langle b|_{X} + H.c%
\label{2.4}%
\end{equation}
where $\omega=\omega_1+\omega_2$ and the recoil wave vector $\mathbf{q}$ is given by the sum of the wave vectors of the beams: $\mathbf{q}=\mathbf{k}_1+\mathbf{k}_2$. The effective Rabi frequency $\Omega=\Omega(\mathbf{r}_X,t)$, assuming the overlap of the laser pulses, provides coupling of the signal sublevels $|b\rangle$ and the Rydberg state $|r\rangle$, see Fig.~\ref{fig1}. Since the laser beams have inhomogeneous spatial profiles it depends on both the atom's position and on time. 

Near the point of the two-photon resonance, where $\omega=\omega_1+\omega_2\sim\omega_{rb}$, the effective Rabi frequency is given by
\begin{equation}
\Omega=-\frac{1}{2}\sum_{n}\frac{\Omega_{rn}^{(2)}\Omega_{nb}^{(1)}}{-\omega_2+\omega_{rn}}%
-\frac{1}{2}\sum_{n}\frac{\Omega_{rn}^{(1)}\Omega_{nb}^{(2)}}{-\omega_1+\omega_{rn}},%
\label{2.5}%
\end{equation}
where $\Omega_{nb}^{(1)}$, $\Omega_{rn}^{(2)}$, \ldots specify the Rabi frequencies of the driving lasers for all the open transitions via the intermediate states $|n\rangle$, and $\omega_{\alpha\beta}$ with $\alpha,\beta=b,n,r\ldots$ denote the transition frequencies, and for the sake of notation simplicity we have omitted in (\ref{2.5}) the dependence on spatial and temporal arguments. Although we use the same notation for the Rabi frequencies for both atoms, the parameters of the exciting pulses are different for each of them. 

The coupling term in (\ref{2.4}) is responsible for the main interaction process leading to the repopulation of atomic states and interference between their spatial motion and spin dynamics. Nevertheless that is an incomplete contribution and the off-resonant laser fields can manifest themselves in the dynamics of the logical states $|a\rangle$ and $|b\rangle$  directly by inducing additional phase shifts within the excitation cycle. These extra shifts can be controlled in an experiment. Since most of our calculations presented below were done under an approximation of nearly rectangular time profiles of the light pulses we can incorporate this kind of correction via ``dressing'' of the original atomic states by adding the energy renormalization terms into the undisturbed Hamiltonian (\ref{2.2}) 
\begin{equation}
\hat{H}_0\to \hat{H}_0+\sum_{\alpha,\beta\ldots}\hbar\left[\Delta_{\alpha}^{(A)}(\mathbf{r}_A)%
+\Delta_{\beta}^{(B)}(\mathbf{r}_B)\right]%
|\alpha,\beta\rangle\langle\alpha,\beta|^{(A,B)},%
\label{2.6}%
\end{equation}
which includes the light shifts $\Delta_{\alpha}^{(A)}=\Delta_{\alpha}^{(A)}(\mathbf{r}_A)$ and $\Delta_{\beta}^{(B)}=\Delta_{\beta}^{(B)}(\mathbf{r}_B)$ to the energy levels enumerated by $\alpha=a,b,r\ldots$ and $\beta=a,b,r\ldots$ for atom $A$ and $B$ respectively. These energy shifts vary with position of the atoms tracing the spatial dependence of the light intensity. 

\subsection{The system dynamics}

\noindent The protocol of Rydberg blockade consists of three subsequent transformation steps and each of the transformations concerns only a particular atom. Therefore let us first describe the dynamics of a single atom in the two-photon excitation process. This dynamic is independent of its proximal neighbor apart from the blockade effect contributed in (\ref{2.1}).

To construct the operator of unitary transformation for a single atom we can simplify the problem and define a single atom wavefunction as
\begin{eqnarray}
|\Psi(t)\rangle &=& \int\frac{d^3p}{(2\pi\hbar)^3}\mathrm{e}^{-\frac{i}{\hbar}\epsilon_{\mathbf{p}}t}%
\left[\mathrm{e}^{-\frac{i}{\hbar}\tilde{\epsilon}_a t}\,c_{a\mathbf{p}}(t)\,|a,\mathbf{p}\rangle\right.%
\nonumber\\%
\nonumber\\%
&&\left.+\mathrm{e}^{-\frac{i}{\hbar}\tilde{\epsilon}_b t}\,c_{b\mathbf{p}}(t)\,|b,\mathbf{p}\rangle
+\mathrm{e}^{-\frac{i}{\hbar}\tilde{\epsilon}_r t}\,c_{r\mathbf{p}}(t)\,|r,\mathbf{p}\rangle\right]%
\nonumber\\%
\label{2.7}%
\end{eqnarray}
where $\epsilon_{\mathbf{p}}=\mathbf{p}^2/2m$ is the kinetic energy of a free atom and the integral expands over its linear momentum $\mathbf{p}$. The basis states are defined in the decoupled representation of the undisturbed Hamiltonian (\ref{2.2}), corrected by the radiation shifts of the energy levels (\ref{2.6}). We have denoted the renormalized internal energy $\tilde{\epsilon}_\alpha=\epsilon_\alpha+\hbar\Delta_{\alpha}(\mathbf{0})$ for any level $\alpha=a,b,r\ldots$, where the light shift is taken at the frame origin coinciding with the focal point of the beam caustic.

In expansion (\ref{2.7}) we have assumed that the atom can occupy three internal states, but only $|b\rangle$ and $|r\rangle$ are involved in the coupled coherent dynamics of the excitation process. The state $|a\rangle$ can be considered as isolated and its probability amplitude $c_{a\mathbf{p}}(t)$ can accumulate a meaningful phase shift during the process, see (\ref{2.6}). The probability amplitudes $c_{r\mathbf{p}}(t)$ and $c_{b\mathbf{p}}(t)$ obey the following coupled dynamics:
\begin{widetext}
\begin{eqnarray}
\dot{c}_{r\mathbf{p}+\hbar\mathbf{q}}&=&
-i\Delta_r(\mathbf{0})\,\mathrm{e}^{\frac{i}{\hbar}\epsilon_{\mathbf{p}+\hbar\mathbf{q}}t}%
\,\left\{\frac{\hbar^2}{z_{R2}^2}\frac{\partial^2}{\partial p_z^2}%
+\frac{2\hbar^2}{w_{02}^2}\triangle_{\bot}+\ldots\right\}%
\mathrm{e}^{-\frac{i}{\hbar}\epsilon_{\mathbf{p}+\hbar\mathbf{q}}t}\,%
c_{r\mathbf{p}+\hbar\mathbf{q}}(t)%
\nonumber\\%
\nonumber\\%
&&+\frac{i}{2}\,\Omega\,\mathrm{e}^{i(\tilde{\omega}_{rb}-\omega)t}\,%
\mathrm{e}^{\frac{i}{\hbar}\epsilon_{\mathbf{p}+\hbar\mathbf{q}}t}%
\,\left\{1+\frac{2\hbar}{z_\ast}\frac{\partial}{\partial p_z}%
+\frac{\hbar^2}{z_\ast^2}\frac{\partial^2}{\partial p_z^2}%
+\frac{2\hbar^2}{w_\ast^2}\triangle_{\bot}+\ldots\right\}\,%
\mathrm{e}^{-\frac{i}{\hbar}\epsilon_{\mathbf{p}}t}%
c_{b\mathbf{p}}(t)%
\nonumber\\%
\nonumber\\%
\nonumber\\%
\dot{c}_{b\mathbf{p}}&=&
-i\Delta_b(\mathbf{0})\,\mathrm{e}^{\frac{i}{\hbar}\epsilon_{\mathbf{p}}t}%
\,\left\{\frac{\hbar^2}{z_{R1}^2}\frac{\partial^2}{\partial p_z^2}%
+\frac{2\hbar^2}{w_{01}^2}\triangle_{\bot}+\ldots\right\}%
\mathrm{e}^{-\frac{i}{\hbar}\epsilon_{\mathbf{p}}t}\,%
c_{b\mathbf{p}}(t)%
\nonumber\\%
\nonumber\\%
&&+\frac{i}{2}\,\Omega^{\ast}\,\mathrm{e}^{-i(\tilde{\omega}_{rb}-\omega)t}\,%
\mathrm{e}^{\frac{i}{\hbar}\epsilon_{\mathbf{p}}t}%
\,\left\{1-\frac{2\hbar}{z_\ast}\frac{\partial}{\partial p_z}%
+\frac{\hbar^2}{z_\ast^2}\frac{\partial^2}{\partial p_z^2}%
+\frac{2\hbar^2}{w_\ast^2}\triangle_{\bot}+\ldots\right\}\,%
\mathrm{e}^{-\frac{i}{\hbar}\epsilon_{\mathbf{p}+\hbar\mathbf{q}}t}%
c_{r\mathbf{p}+\hbar\mathbf{q}}(t).%
\label{2.8}%
\end{eqnarray}
\end{widetext}
Here $\tilde{\omega}_{\alpha\beta}$ with $\alpha,\beta=a,b,r\ldots$ denote the dressed transition frequencies; $w_{0j}$ and $z_{Rj}$ with $j=1,2$ are respectively the beam waists and Rayleigh ranges of the Gaussian laser beams, and we have defined the set of effective parameters:
\begin{eqnarray}
\frac{2}{w_{\ast}^2}&\equiv&\frac{1}{w_{01}^2}+\frac{1}{w_{02}^2},%
\nonumber\\%
\frac{2}{z_{\ast}}&\equiv&\frac{1}{z_{R1}}+\frac{1}{z_{R2}},%
\nonumber\\%
\frac{2}{z_{\ast}^2}&\equiv&\frac{1}{z_{R1}^2}+\frac{1}{z_{R2}^2},%
\label{2.9}%
\end{eqnarray}
where the conventional Gaussian beam parameters in the right-hand side are defined in Appendix \ref{Appendix_A}.

These equations are presented for the pulses shaped by rectangular profiles having the same duration, such that all the Rabi frequencies are supposed to be constant during the pulses. We have estimated the light shifts at the origin point, associated with the focal point having the maximal light intensity, as
\begin{eqnarray}
\Delta_b(\mathbf{0})&\simeq&\frac{1}{4}\sum_{n}\frac{|\Omega_{nb}^{(1)}(\mathbf{0})|^2}%
{\omega_1-\omega_{nb}}%
\nonumber\\%
\Delta_r(\mathbf{0})&\simeq&-\frac{1}{4}\sum_{n}\frac{|\Omega_{rn}^{(2)}(\mathbf{0})|^2}%
{\omega_2-\omega_{rn}}%
\label{2.10}%
\end{eqnarray}
and have kept the main contributions with respect to the relatively small detuning $\omega_2-\omega_{rn}\sim\omega_{nb}-\omega_1$.

The key feature of equations (\ref{2.8}) is the presence of differential terms containing the first and second order partial derivatives and the transverse Laplace operator $\triangle_{\bot}$ acting on the linear momentum arguments. These terms have resulted from the expansion (\ref{a.7}) for the field amplitude in the vicinity of the focal point. Once we approximate the Gaussian mode for both of the beams by an infinite plane wave, we arrive at the textbook result, i.e. to the coupled equations describing the coherent dynamics in a two-level system:
\begin{eqnarray}
\lefteqn{\dot{c}_{r\mathbf{p}+\hbar\mathbf{q}}\!=\!\frac{i}{2}\Omega%
\exp\left[i\!\left(\tilde{\omega}_{rb}+\frac{\mathbf{q}\!\cdot\!\mathbf{p}}{m}%
+\frac{\hbar\mathbf{q}^2}{2m}-\omega\right)\!t\right]%
c_{b\mathbf{p}}(t)}%
\nonumber\\%
\nonumber\\%
&&\dot{c}_{b\mathbf{p}}\!=\!\frac{i}{2}\Omega^{\ast}%
\exp\left[-i\!\left(\tilde{\omega}_{rb}+\frac{\mathbf{q}\!\cdot\!\mathbf{p}}{m}%
+\frac{\hbar\mathbf{q}^2}{2m}-\omega\right)\!t\right]%
c_{r\mathbf{p}+\hbar\mathbf{q}}(t).%
\nonumber\\%
\label{2.11}%
\end{eqnarray}
The added terms, having an evident signature of the diffusion process, correct the system dynamics towards parametric heating and dephasing of the internal state of the atom during its excitation. Physically, these terms reveal an uncertainty in the momentum conservation associated with the random drift of the atom through the inhomogeneous field profile.

Assume that at an initial moment of time $t=0$ the probability amplitudes in expansion (\ref{2.7}) had the given values $c_{a\mathbf{p}}(0)$, $c_{b\mathbf{p}}(0)$, and $c_{r\mathbf{p}+\hbar\mathbf{q}}(0)$. Then after a light pulse of duration $\tau$ in accordance with equations (\ref{2.8}) the amplitudes would transform to
\begin{widetext}
\begin{equation}
\left(\begin{array}{c}c_{r\mathbf{p}+\hbar\mathbf{q}}(\tau) \\ c_{b\mathbf{p}}(\tau) \\ c_{a\mathbf{p}}(\tau) \end{array} \right)
=\left(\begin{array}{ccc} \hat{U}_{rr}(\tau)&\hat{U}_{rb}(\tau)& 0 \\
\hat{U}_{br}(\tau)&\hat{U}_{bb}(\tau)&0\\
0&0& 1
\end{array}\right)
\left(\begin{array}{c}c_{r\mathbf{p}+\hbar\mathbf{q}}(0) \\ c_{b\mathbf{p}}(0) \\ c_{a\mathbf{p}}(0) \end{array} \right),
\label{2.12}
\end{equation}
\end{widetext}
where each element in the first two rows of the transformation matrix is an integral-type operator acting on the momentum variables. As commented above (see definition of the undisturbed Hamiltonian (\ref{2.6})) the probability amplitude for all the basis states accumulate additional phases associated with the light shifts induced by the driving fields. Here in (\ref{2.12}) these phases are incorporated into the respective shifts of the renormalized energy levels in the definition of the wavepacket (\ref{2.7}), expanded in the basis of dressed states. 

If we reduce (\ref{2.8}) to its lighter form (\ref{2.11}) the transformation would contain the c-number matrix elements such that the solution (\ref{2.12}) would reveal the well known periodic time beats in the two-level system with subsequent occupations by the atom of either state $|b,\mathbf{p}\rangle$ or $|r,\mathbf{p}+\hbar\mathbf{q}\rangle$.

\subsection{The entanglement protocol}\label{Section_IIC}
\noindent The entanglement protocol is described as the following transformation of the originally disentangled atomic state:
\begin{equation}
|\psi\rangle_{AB}=\hat{\cal U}_3\,\hat{\cal U}_2\,\hat{\cal U}_1|\psi\rangle_{A}|\psi\rangle_{B},%
\label{2.13}%
\end{equation}
where at the first step
\begin{equation}
\hat{\cal U}_1=\hat{U}^{(\pi)}_A\otimes\hat{I}_B%
\label{2.14}%
\end{equation}
with the operator $\hat{U}^{(\pi)}_A$ defined by Eq.~(\ref{2.12}) and acting on the control atom during the time $|\Omega|\tau=\pi$. The state of the target atom is unchanged, which is formally expressed by the unit operator $\hat{I}_B$.

At the second step we involve the cooperative bus-type interaction; the action on the target atom $B$ which is sensitive to the state of the control atom $A$. The transformation matrix $\hat{\cal U}_2$ is given by
\begin{eqnarray}
\hat{\cal U}_2&=&\big{[}|a\rangle\langle a|_{A}+|b\rangle\langle b|_{A}\big{]}\otimes%
\hat{U}^{(2\pi)}_B
\nonumber\\%
\nonumber\\%
&&+|r\rangle\langle r|_{A}\otimes\left.\hat{U}^{(2\pi)}_B\right|_{\delta_R},%
\label{2.15}%
\end{eqnarray}
where in both terms the transformation (\ref{2.12}) is applied for the duration $|\Omega|\tau=2\pi$, but in the second term we have assumed the shifted transition frequency $\tilde{\omega}_{rb}\to\tilde{\omega}_{rb}+\delta_R$, see (\ref{2.1}). 

At the third step of the protocol we apply the same transformation as at the first step, such that $\hat{\cal U}_3=\hat{\cal U}_1$. That returns the atom $A$ to the ground state. The final state cannot in general be expressed as the product of independent wavefunctions and accumulates the quantum correlations. 

Let us naively assume that in an ideal scenario the subsequent action of three operators in (\ref{2.13}) ignores the momentum variables and concerns only the internal ground spin states $|a\rangle$ and $|b\rangle$ of both the atoms. Then we are allowed to manipulate only with the spin subsystem of both the atoms. Thus if (with making use of the interaction representation with respect to the internal Hamiltonian (\ref{2.2})) we have originally prepared the states
\begin{eqnarray}
|\psi\rangle_{A}&=&\frac{1}{\sqrt{2}}\left[|a\rangle+|b\rangle\right]_A%
\nonumber\\%
|\psi\rangle_{B}&=&\frac{1}{\sqrt{2}}\left[|a\rangle+|b\rangle\right]_B.%
\label{2.16}%
\end{eqnarray}
Then we arrive at
\begin{equation}
|\psi\rangle_{AB}=\frac{1}{2}\left[|a,a\rangle-|b,a\rangle-|a,b\rangle-|b,b\rangle\right]_{AB},%
\label{2.17}%
\end{equation}
which corresponds to an application of a CZ quantum logic gate in the system of two spin qubits. Nevertheless, as is clear from the discussion above that this can be only approximately done and below, via a simple estimate, we explain why such an ideal scenario fails and show the critical benchmark for a potentially attainable fidelity for the state (\ref{2.17}).

\subsection{Importance of the recoil effect}\label{Section_IID}

\noindent Let us consider the blockade scheme and track the linear momentum transfer accompanying the two-photon process under conditions maximally approaching an ideal scenario. Assume that originally both atoms are cooled to the ground states of the respective trap wells. Then after an excitation cycle their vibrational motion would be activated because of the linear momentum transfer in the two-photon excitation process, see \cite{Saffman_PRL2019,Saffman_PRA2021}. That produces a coherent mode parameterized by a displacement depending on the time spent by each atom in the upper Rydberg state. Both atoms, when de-excited to the ground states, will approximately evolve to motional coherent states, i.e. will have the same shape of the wave-packet profile periodically oscillating in the trap well. 

Let us denote the ground vibration state as $|0\rangle$ and the coherent states as $|\boldsymbol{\alpha}\rangle$ and $|\boldsymbol{\alpha}'\rangle$ for the atoms $A$ and $B$ respectively, treating the coherent amplitudes as vector quantities $\boldsymbol{\alpha}=\alpha_x,\alpha_y,\alpha_z$ and $\boldsymbol{\alpha}'=\alpha'_x,\alpha'_y,\alpha'_z$. Instead of (\ref{2.17}) we can expect the following final state
\begin{eqnarray}
|\psi\rangle_{AB}^{(s+vib)}&=&\frac{1}{2}\left[|a,0;a,0\rangle-|b,\boldsymbol{\alpha};a,0\rangle\right.%
\nonumber\\%
&&\left.-|a,0;b,\boldsymbol{\alpha}'\rangle%
-|b,\boldsymbol{\alpha};b,\boldsymbol{\alpha}'\rangle\right]_{AB}%
\label{2.18}%
\end{eqnarray}
which is an entangled state with respect to both the spin states and vibrational modes.

There is no experimental resource to control vibrational motion and we should convert the state (\ref{2.18}) to the mixed spin state described by the density matrix
\begin{equation}
\hat{\rho}^{(s)}=\mathrm{Tr}'_{vib}\;|\psi\rangle\langle\psi|_{AB}^{(s+vib)}%
\label{2.19}%
\end{equation}
Then we can estimate fidelity ${\cal F}$ of reproduction of the ideal state (\ref{2.17}) as
\begin{eqnarray}
{\cal F}&=&\langle\psi|\hat{\rho}^{(s)}|\psi\rangle_{AB}%
\nonumber\\%
\nonumber\\%
&=&\frac{1}{4}\left[1+\mathrm{e}^{-|\boldsymbol{\alpha}|^2/2}%
 +\mathrm{e}^{-|\boldsymbol{\alpha}'|^2/2}%
 +\mathrm{e}^{-(|\boldsymbol{\alpha}|^2+|\boldsymbol{\alpha}'|^2)/2}\right],%
 \nonumber\\%
 \label{2.20}%
\end{eqnarray}
where the displacement parameters $\boldsymbol{\alpha}$ and $\boldsymbol{\alpha}'$ depend on the linear momentum transfer $\hbar\mathbf{q}$ and on the duration of the process. 

In theory, the problem with optimization of (\ref{2.20}) could be resolved by two-photon excitation under quasi-degenerate conditions i.e. by choosing $\omega_1\sim\omega_2$ and $\mathbf{q}=\mathbf{k}_1+\mathbf{k}_2\sim\mathbf{0}$. Evidently in such a case we should exclude the precise equality to avoid that the atom could be excited by two photons derived from the same laser beam that is split into two counter-propagating beams. In practice, however it would not be so easy to stabilize the controllable excitation in a far off-resonant frequency domain for the intermediate atomic states. Thus one needs to stay close to the resonance of some intermediate state, which makes the condition $\omega_1\sim\omega_2$ inaccessible for the existing experimental capabilities. In reality we are limited within a few percents for $\alpha$ and $\alpha'$ and the attained fidelity can be upper bounded by ${\cal F}<0.9995$ at best.


\section{Incoherent losses}\label{Section_III}

\begin{figure}[tp]
\includegraphics[width=8.6cm]{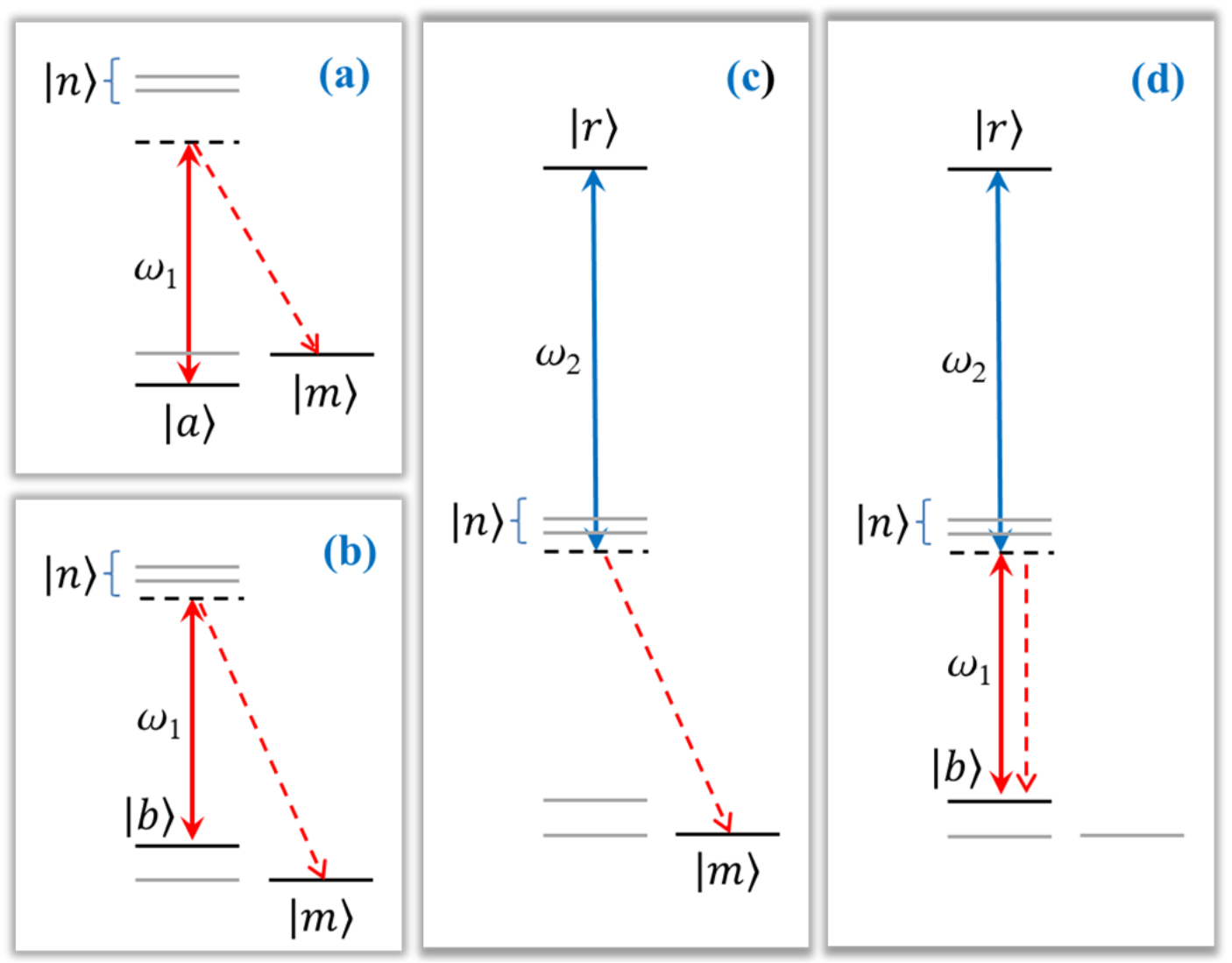}
\caption{Transition diagrams for different channels of the incoherent losses: (a),(b) spontaneous Raman scattering of the field mode $\omega_1$ from the qubit states $|a\rangle$ and $|b\rangle$ to a Zeeman state $|m\rangle$; (c) same for mode $\omega_2$ scattered from the Rydberg level $|r\rangle$; (d) leakage from the two-photon coherent excitation channel to the CPT ``dark'' state, see the text.}
\label{fig2}%
\end{figure}

\noindent The dynamical description of the physical processes accompanying the entanglement protocol presented in the preceding section is incomplete since the entire system is open for interaction with the environment. The coherent modes $\omega_1$ and $\omega_2$ initiate the processes of incoherent Rayleigh and Raman scattering via intermediate states, as it is symbolically clarified by the diagrams in Fig.~\ref{fig2}. For dynamics of a single qubit the Rayleigh scattering preserves the atomic coherence, but the Raman scattering induces irreversible loss of coherence, see \cite{Meschede2005,SingleQubit2021}. Here the situation is more complicated and the different output scattering channels, initiated by different modes, overlap and interfere each over. During a short protocol time the losses are weak and in our further estimates we will simplify the model and completely ignore any kinematic manifestations, considering the atoms as immobile scatterers. Even under such assumptions the conventional simulations guided by a rate-type master equation would be problematic if the time duration of the driving pulses were comparable or less than the natural lifetime $\gamma^{-1}$ of the intermediate $D_{1/2}$ state. In the latter case, the transient dynamics induces the fast non-vanishing Rabi oscillations to the process. That would make small but meaningful corrections to our further estimates purposing in effective processing of the quantum correlations. The transient dynamics becomes negligible for duration sufficiently longer than $\gamma^{-1}$, which we will additionally assume here as a condition typically fulfilled in an experiment.

\subsection{The scattering tensor and transition rates}

\noindent The spontaneous scattering can be conveniently framed by the formalism of the scattering tensor, see \cite{Plachek,BERESTETSKII19821}. Under the rotating wave approximation (RWA) this tensor coincides with the dynamical polarizability tensor. The specifics of the considered situation is that the atom scattering the light can originally either occupy one of the low energy qubit states $|a\rangle$ or $|b\rangle$ or be excited in the Rydberg state $|r\rangle$. 

Then for the scattering from state $|\alpha\rangle=|a\rangle,\,|b\rangle$ it is given by
\begin{equation}
\alpha_{ik}^{(m\alpha)}=-\frac{1}{\hbar}\sum_{n}\frac{(d_i)_{mn}(d_k)_{n\alpha}}{\omega_1-\omega_{n\alpha}+i0}.%
\label{3.1}%
\end{equation}
But for the scattering from the upper state $|r\rangle$ there is a difference in the structure of the denominator:
\begin{equation}
\alpha_{ik}^{(mr)}=-\frac{1}{\hbar}\sum_{n}\frac{(d_i)_{mn}(d_k)_{nr}}{-\omega_2+\omega_{rn}+i0}.%
\label{3.2}%
\end{equation}
In these equations the tensor indices $i,k=x,y,z$ enumerate the vector components of the transition dipole moments, $|m\rangle$ specify those quantum states which are repopulated after the scattering event. 

Here the issue arises whether it is correct to understand the process initiated from the upper Rydberg level and described by the amplitude (\ref{3.2}) as scattering, when instead of annihilation the virtual transition to the intermediate level is assisted by creation of the photon in the incident mode $\omega_2$. That seems to contradict the conventional definition of the scattering phenomenon and, strictly speaking, should be designated as a two-photon emission. But we can point out that for the processes shown in Fig.~\ref{fig2} there is actually neither annihilation nor creation of a photon in the driving modes $\omega_1$ and $\omega_2$. Both modes exist in the coherent state, the amplitudes of which are insensitive to either attenuation or amplification at the single photon level. In other words, under the discussed conditions both light modes are unchanged but able to stimulate the emission of an off-resonant photon in any direction with a certain probability and to repopulate the atom randomly into any accessible Zeeman sublevel in its ground state. Regardless of whether the atom occupies $|a\rangle$, $|b\rangle$ or $|r\rangle$ states both channels work similarly and for the sake of physical clarity we call both of them ``incoherent scattering.'' 

The scattering processes lead to irreversible losses and to the purity reduction of the entangled spin state shared by the atoms. The transition rates for the scattering processes, considered independently for each mode, can be expressed by the Rabi frequencies of the driving lasers and are given by
\begin{equation}
w_{\alpha\to m}=\frac{{\omega'}^3}{8\pi\hbar c^3}\int\!do'\left|\sum_{n}\frac{(\mathbf{e}'\cdot\mathbf{d})_{mn}\,\Omega_{n\alpha}^{(1)}}{\omega_1-\omega_{n\alpha}+i0}\right|^2,%
\label{3.3}%
\end{equation}
with $\alpha=a,b$ and
\begin{equation}
w_{r\to m}=\frac{{\omega'}^3}{8\pi\hbar c^3}\int\!do'\left|\sum_{n}\frac{(\mathbf{e}'\cdot\mathbf{d})_{mn}\,\Omega_{rn}^{(2)\ast}}{-\omega_2+\omega_{rn}+i0}\right|^2,%
\label{3.4}%
\end{equation}
where $\omega'=\omega_1+\omega_{\alpha m}$ in (\ref{3.3}) and $\omega'=\omega_{rm}-\omega_2$ in (\ref{3.4}) is the frequency of the emitted photon and $\mathbf{e}'$ is its polarization. The integral is evaluated over the solid angle $do'$ for all the scattering directions. However the complete description of the processes visualized by diagrams (a)-(c) in Fig.~\ref{fig2} incorporates various interference contributions, provided by conditions of the two-photon resonance, which we further clarify.

\subsection{Evolution of the density matrix}\label{Section_IIIB}
\noindent The spatial degrees of freedom are inaccessible for direct detection and in reality we deal with an open system, which is relevantly described by the reduced two-particle density matrix, defined for the collective internal spin state of the control and target atoms: $\rho_{\alpha',\beta';\alpha,\beta}(t)$, where $\alpha,\alpha'$ and $\beta,\beta'$ enumerate the basis quantum states belonging to atoms $A$ and $B$ respectively. Evolution of this matrix obeys the master equation, which includes both the dynamical transformation and the irreversible relaxation processes. The latter lead to entanglement losses and are mainly connected with the incoherent scattering channels described above which evolve the system to a statistically mixed state. To qualify the different contributions of the spontaneous scattering to the complete evolution of the density matrix compiled from the elementary processes shown in Fig.~\ref{fig2} (a)-(c), we will follow the Keldysh's diagram method, and the supporting graph images are introduced in Appendix \ref{Appendix_C}.

In order to unify our discussion, which is in what follows based on the density matrix formalism, with the dynamical description presented in the preceding section, we will follow the interaction representation, eliminating free energy oscillations induced by the internal Hamiltonian (\ref{2.2}), (\ref{2.6}) throughout our derivation below. We neglect the spatial motion of atoms and assume the conditions of the exact two-photon resonance between the dressed working states, such that $\omega_1+\omega_2=\tilde{\omega}_{rb}$, but we admit that other Zeeman states can have shifted energies because of the magnetic field, anisotropy of light shifts, etc.  

Let us consider the scattering of each mode on the control atom $A$ as represented by diagrams (\ref{c.5}) and (\ref{c.6}). If the control atom occupies one of the qubit states $|\alpha\rangle=|a\rangle,\,|b\rangle$ and the target atom $B$ occupies any accessible state $|\beta\rangle$, $|\beta'\rangle$ we add the term
\begin{equation}
\dot{\rho}_{\alpha,\beta';\alpha,\beta}=\ldots -w_{\alpha}\;\rho_{\alpha,\beta';\alpha,\beta}(t)%
\label{3.5}%
\end{equation}
with
\begin{equation}
w_{\alpha}=\sum_{m}w_{\alpha\to m}=\gamma\sum_{n}\frac{|\Omega_{n\alpha}^{(1)}|^2}{4\Delta_{n\alpha}^2},%
\label{3.6}%
\end{equation}
where $\gamma$ is the natural radiative decay rate of the state $|n\rangle$, belonging to the intermediate levels, which is assumed to be independent on $|n\rangle$, and $\Delta_{n\alpha}=\omega_1-\omega_{n\alpha}$. We keep the interaction of the mode $\omega_1$ with both qubit states in our model, but the depopulation rate $w_a$ is essentially smaller than $w_b$ due to higher detuning. In the right-hand sides of (\ref{3.5}) and in all other equations appearing below in this section we indicate the omitted terms contributing to the total time derivatives of the particular components of the density matrix by ellipses.

Similarly if the control atom $A$ occupies the Rydberg state $|\alpha\rangle=|r\rangle$ we add the term
\begin{equation}
\dot{\rho}_{r,\beta';r,\beta}=\ldots -w_r\;\rho_{r,\beta';r,\beta}(t)%
\label{3.7}%
\end{equation}
with
\begin{equation}
w_r=\sum_{m}w_{r\to m}=\gamma\sum_{n}\frac{|\Omega_{rn}^{(2)}|^2}{4\Delta_{rn}^2},%
\label{3.8}%
\end{equation}
where $\Delta_{rn}=\omega_2-\omega_{rn}$. 

For coherences between $|r\rangle$ and $|b\rangle$, between $|b\rangle$ and $|a\rangle$, and between $|r\rangle$ and $|a\rangle$ we subsequently obtain
\begin{eqnarray}
\dot{\rho}_{r,\beta';b,\beta}&=&\ldots -\left[\frac{w_r}{2} + \frac{w_b}{2}\right]\,\rho_{r,\beta';b,\beta}(t),%
\nonumber\\%
\dot{\rho}_{b,\beta';a,\beta}&=&\ldots -\left[\frac{w_b}{2} + \frac{w_a}{2}\right]\,\rho_{b,\beta';a,\beta}(t),%
\nonumber\\%
\dot{\rho}_{r,\beta';a,\beta}&=&\ldots -\left[\frac{w_r}{2} + \frac{w_a}{2}\right]\,\rho_{r,\beta';a,\beta}(t).%
\label{3.9}%
\end{eqnarray}
The coherence between $|r\rangle$ and $|a\rangle$ is created from the original coherence between $|b\rangle$ and $|a\rangle$ by excitation of the atom from $|b\rangle$ to $|r\rangle$ by the $\pi$-pulse. The same terms have to be added for the Hermitian conjugated components. Note that the terms (\ref{3.5}), (\ref{3.7}), and (\ref{3.9}) are activated only on the stages of $\pi$-pulses, see Fig.~\ref{fig1} and at these stages there is no interaction of the driving fields with atom $B$, which as a spectator is allowed to occupy the ground state only, such that $\beta,\beta'\neq r$ in all these equations. 

If the target atom $B$ is excited by a $2\pi$-pulse the situation is somewhat different. At this stage, for diagonal components of the density matrix we obtain 
\begin{equation}
\dot{\rho}_{\alpha',\beta;\alpha,\beta}=\ldots -w_\beta\;\rho_{\alpha',\beta;\alpha,\beta}(t)%
\label{3.10}%
\end{equation}
if $|\beta\rangle=|a\rangle,\,|b\rangle$ with $w_{\beta}$ given by (\ref{3.6}) with $\alpha\to\beta$, and 
\begin{equation}
\dot{\rho}_{\alpha',r;\alpha,r}=\ldots -w_r\;\rho_{\alpha',r;\alpha,r}(t)%
\label{3.11}%
\end{equation}
with $w_{r}$ given by (\ref{3.8}). For coherences we get
\begin{eqnarray}
\dot{\rho}_{\alpha',r;\alpha,b}&=&\ldots -\left[\frac{w_r}{2} + \frac{w_b}{2}\right]\,\rho_{\alpha',r;\alpha,b}(t)%
\nonumber\\%
\dot{\rho}_{\alpha',b;\alpha,a}&=&\ldots -\left[\frac{w_b}{2} + \frac{w_a}{2}\right]\,\rho_{\alpha',b;\alpha,a}(t)%
\nonumber\\%
\dot{\rho}_{\alpha',r;\alpha,a}&=&\ldots -\left[\frac{w_r}{2} + \frac{w_a}{2}\right]\,\rho_{\alpha',r;\alpha,a}(t)%
\label{3.12}%
\end{eqnarray}
The blockade effect prevents both atoms from simultaneously populating the Rydberg states, such that in (\ref{3.11})  $\alpha,\alpha'\neq r$ and in the first and the third lines of (\ref{3.12}) $\alpha'\neq r$, but in (\ref{3.10}) and in the second line of (\ref{3.12}) the control atom can occupy any state. Also note that, within the accuracy of our consideration for such correcting terms, if the atom $A$ is not in $|r\rangle$ it can only occupy the state $|a\rangle$, since $|b\rangle$ is depopulated in accordance with the protocol.

The contributions considered above correct the dynamics of those components of the density matrix which are generated by the driving fields as described in the preceding section. These terms are responsible for depopulation of both the atoms away from the working levels, and lead to decoherence of the two-particle density matrix, and, as a consequence, to reduction of quantum entanglement. However there is a set of parallel processes, which recover the population balance among the Zeeman sublevels of the ground state and affect the quantum correlations as well.
\\
\subsubsection{The repopulation of atoms by optical pumping}

\noindent The loss of atomic population from the working levels is partly compensated for by the opposite process of repopulation by optical pumping induced by incoherent scattering, and providing the atomic polarization transfer to the ground state. Here, as optical pumping, we refer to those channels, which are developing independently for each mode and insensitively to the conditions of two-photon resonance, and which are represented by diagrams (\ref{c.7}) and (\ref{c.8}) in Appendix \ref{Appendix_C}.  

Consider the control atom $A$. Transformation of the density matrix due to optical pumping is described by the following income-type term
\begin{widetext}
\begin{eqnarray}
\dot{\rho}_{m',\beta';m,\beta}(t)&=&\ldots+\gamma\,\sum_{\alpha',\alpha}\exp\left[i(\tilde{\omega}_{m'm}-\tilde{\omega}_{\alpha'\alpha})t\right]\,\rho_{\alpha',\beta';\alpha,\beta}(t)%
\sum_{n',n}\frac{\Omega_{n'\alpha'}^{(\alpha')}\,\Omega_{n\alpha}^{(\alpha)\ast}}{4\Delta_{n'\alpha'}\,\Delta_{n\alpha}}%
\sum_{q}C_{F'_0M'_0\, 1q}^{F'M'}C_{F_0M_0\, 1q}^{FM}%
\nonumber\\%
&&\times\; (-)^{F_0-F'_0}\left[(2F'_0+1)(2F_0+1)\right]^{1/2}(2J+1)%
\;\left\{\begin{array}{ccc} S & I & F'_0\\ F' & 1 & J \end{array}\right\}%
\left\{\begin{array}{ccc} S & I & F_0\\ F & 1 & J \end{array}\right\}%
\label{3.13}%
\end{eqnarray}
\end{widetext}
where $m=F_0,M_0$, $m'=F'_0,M'_0$ are the repopulated states belonging to the ground manifold and expressed in the basis of the coupled electron and nuclear spin angular momenta.  We assume the working set $|a\rangle,\,|b\rangle,\,|r\rangle$ and possible coherent superposition between $|a\rangle$ and $|b\rangle$ as the depopulated states $|\alpha\rangle,\,|\alpha'\rangle$ contributing to the right-hand side of this equation. The intermediate states $|n\rangle$ and $|n'\rangle$ coherently coupled with them via the driving fields are specified by the total angular momenta, given by the sum of the electron orbital and spin momenta with the nuclear spin: $n=F,M$; $n'=F',M'$. Other quantum numbers specifying these states are $S=1/2$, $I$ and $J$ -- the electron spin, nuclear spin, and total electron angular momentum, respectively. The total electron angular momentum $J$ is assumed to be the same for any of the states $|n\rangle,\,|n'\rangle$ . We have superscribed here the Rabi frequencies with the same index as the respective working level for the sake of notation convenience. The transition matrix is expressed by the angular momentum functions, namely by Clebsch-Gordon coefficients $C^{\dots}_{\ldots\,\ldots}$ and 6j-symbols $\{\ldots\}$, see \cite{LaLfIII,varshalovich}.

The repopulating term (\ref{3.13}), in particular, can provide the resonant transfer of the ground state hyperfine coherence, which originally exists for the superpositions of the qubit levels $|a\rangle$ and $|b\rangle$. As a result of repopulation the spin coherence between other Zeeman states belonging to different hyperfine sublevels may be created. So the system would have the internal quantum correlations but lose entanglement shared by the logical states.

Optical pumping initiating the polarization transfer for the target atom $B$ can be similarly described with an appropriate change of the quantum numbers in (\ref{3.13}). An exception is the case when the control atom occupies the Rydberg state and atom $B$ is not allowed to occupy the state $|r\rangle$ since the double occupation $|r,r\rangle$ is prevented by the blockade effect. It might seem that we could transfer some specific terms containing coherence for atom $A$ being superposed between states $|r\rangle$ and $|a\rangle$ via the optical pumping channel. Such a coherence can be preliminarily created from the qubit state (\ref{2.16}) by a $\pi$-pulse conversion of its $|b\rangle$-part to $|r\rangle$. It might be suggested that the particular matrix element $\rho_{a,r;r,\beta}$ (with $\beta=a,b$) and its Hermitian conjugate $\rho_{r,\beta;a,r}$ can be involved in the repopulation process while atom $B$ is excited by a $2\pi$-pulse. Their depopulation components indeed exist and were already taken into consideration regarding the depopulating process in the first and third lines of (\ref{3.12}). However we cannot construct their optical pumping repopulation term since it can provide transfer only to the ground state, see diagrams (a)-(c) in Fig.~\ref{fig2}. So the suggested repopulating process to states $|r,\beta\rangle$ would be extremely off-resonant, which is formally expressed by highly oscillating term in the right-hand side of (\ref{3.13}), and actually would be far beyond all of the approximations made.

\subsubsection{Coherent population trapping}

\noindent The simultaneous excitation of a three level transition with two field modes can under certain conditions transform the system behavior to a manifestation of a coherent population trapping (CPT) phenomenon, see \cite{Arimondo1976,Scully}. Any element in the unitary subspace formed by linear span of two metastable states, in our case $|b\rangle$ and $|r\rangle$, can be alternatively decomposed into two orthogonal superpositions of the so called ``bright'' and ``dark'' states.  Being excited by two coherent modes and approaching the steady state regime the system will eventually leak to the dark state, which eliminates its further interaction with the driving fields, see Fig.~\ref{fig2}(d). In the considered configuration, assuming a pulsed excitation, we only deal with the seeding stage of this process partly disentangling the qubits.

The spontaneous leakage to the CPT dark state can be foreseen from the non-Hermitian correction to the interaction part of the effective Hamiltonian (\ref{2.4}) and (\ref{2.5}). Indeed in a rigorous approach, the additional terms beyond the effective Hamiltonian concept arise from the weak relaxation of the optical coherence, assisting the two-photon excitation during very short virtual transition time $\Delta_{nb}^{-1}$ at the rate of $\gamma/2$, see Fig.~\ref{fig2}(d). The process is expressed by diagrams (\ref{c.9}) and (\ref{c.10}) in Appendix \ref{Appendix_C}. Note that the disparities between the modes $\omega_1$ and $\omega_2$ are crucially important for a fair observation of the CPT resonance, so for consistency we have to leave only the first term in the structure of the effective Hamiltonian (\ref{2.5}). In the case of the control atom $A$ being excited we obtain
\begin{widetext}
\begin{eqnarray}
\dot{\rho}_{b,\beta';b,\beta}&=&\ldots -\frac{\gamma}{2}\sum_{n}\left[\frac{\Omega_{rn}^{(2)\ast}\Omega_{nb}^{(1)\ast}}{4\Delta_{nb}^2}\,\rho_{r,\beta';b,\beta}(t)%
+\frac{\Omega_{rn}^{(2)}\Omega_{nb}^{(1)}}{4\Delta_{nb}^2}\,\rho_{b,\beta';r,\beta}(t)\right]%
\nonumber\\%
\nonumber\\%
\dot{\rho}_{r,\beta';r,\beta}&=&\ldots -\frac{\gamma}{2}\sum_{n}\left[\frac{\Omega_{rn}^{(2)\ast}\Omega_{nb}^{(1)\ast}}{4\Delta_{nb}^2}\,\rho_{r,\beta';b,\beta}(t)%
+\frac{\Omega_{rn}^{(2)}\Omega_{nb}^{(1)}}{4\Delta_{nb}^2}\,\rho_{b,\beta';r,\beta}(t)\right]%
\nonumber\\%
\nonumber\\%
\dot{\rho}_{r,\beta';b,\beta}&=&\ldots -\frac{\gamma}{2}\sum_{n}\frac{\Omega_{rn}^{(2)}\Omega_{nb}^{(1)}}{4\Delta_{nb}^2}\;\left[\phantom{\frac{}{}}\rho_{b,\beta';b,\beta}(t)+\rho_{r,\beta';r,\beta}(t)\right]%
\nonumber\\%
\dot{\rho}_{b,\beta';r,\beta}&=&\ldots -\frac{\gamma}{2}\sum_{n}\frac{\Omega_{rn}^{(2)\ast}\Omega_{nb}^{(1)\ast}}{4\Delta_{nb}^2}\;\left[\phantom{\frac{}{}}\rho_{b,\beta';b,\beta}(t)+\rho_{r,\beta';r,\beta}(t)\right].%
\label{3.14}%
\end{eqnarray}
\end{widetext}
Here we have assumed that the atom $B$ exists in its ground state such that $\beta,\beta'\neq r$. These terms indicate leakage from the coherent two-photon excitation dynamics and spontaneous transition of the atom to any accessible Zeeman sublevel of the ground state. Then, as shown by diagram (d) in Fig.~\ref{fig2}, it can with certain probability spontaneously populate the state $|b\rangle$ in this process, recovering its coherent coupling to the state $|r\rangle$ (bright state) or eliminating the two-photon interaction (dark state). After a round of such spontaneous cycles, the atom will eventually transit to the dark state.

Furthermore, recall that a key element of the entanglement protocol is the dynamical transformation of coherence between $|b\rangle$ and $|a\rangle$ to coherence between $|r\rangle$ and $|a\rangle$. It primarily follows the dynamical evolution but, as expressed by (\ref{3.9}), is partly distorted by weak relaxation induced by the incoherent scattering of each mode, which is treated in (\ref{3.9}) as happening independently from either energy level. Here we revise this point and extend the decoherence process by an option to spontaneously emit a photon in free space fulfilling the two photon resonance coupling of $|r\rangle$ and $|b\rangle$. That is also represented by diagrams (\ref{c.9}) and (\ref{c.10}) and expressed by the following corrections to the relaxation of these coherences:

\begin{eqnarray}
\dot{\rho}_{b,\beta';a,\beta}&=&\ldots -\frac{\gamma}{2}\sum_{n}\frac{\Omega_{rn}^{(2)\ast}\Omega_{nb}^{(1)\ast}}{4\Delta_{nb}^2}\,\rho_{r,\beta';a,\beta}(t)%
\nonumber\\%
\nonumber\\%
\dot{\rho}_{r,\beta';a,\beta}&=&\ldots -\frac{\gamma}{2}\sum_{n}\frac{\Omega_{rn}^{(2)}\Omega_{nb}^{(1)}}{4\Delta_{nb}^2}\,\rho_{b,\beta';a,\beta}(t).%
\nonumber\\
\label{3.15}%
\end{eqnarray}
where the free precession of the spin coherence at frequency $\tilde{\omega}_{ba}$ combines with an external excitation induced by the driving fields such that $\tilde{\omega}_{ra}=\tilde{\omega}_{ba}+\omega_1+\omega_2$. 

The leakage of the atom from the working channel in the two-photon excitation process, expressed by (\ref{3.14}) and (\ref{3.15}), is balanced by the backward repopulation process, which in case of atom $A$ is given by the following incoming-type term:
\begin{widetext}
\begin{eqnarray}
\dot{\rho}_{m',\beta';m,\beta}(t)&=&\ldots+\gamma\,\exp\left[i\tilde{\omega}_{m'm}t\right]\,\sum_{\alpha=a,b}\;%
\sum_{n',n}\left[\exp\left[-i\tilde{\omega}_{b\alpha}t\right]\frac{\Omega_{rn'}^{(2)\ast}\Omega_{n\alpha}^{(1)\ast}}{4(-\Delta_{rn'})\Delta_{n\alpha}}\,\rho_{r,\beta';\alpha,\beta}(t)\right.%
\nonumber\\%
&&\left.+\,\exp\left[-i\tilde{\omega}_{\alpha b}t\right]\frac{\Omega_{rn}^{(2)}\Omega_{n'\alpha}^{(1)}}{4(-\Delta_{rn})\Delta_{n'\alpha}}\,\rho_{\alpha,\beta';r,\beta}(t)\right]\sum_{q}C_{F'_0M'_0\, 1q}^{F'M'}C_{F_0M_0\, 1q}^{FM}%
\nonumber\\%
&&\times\;(-)^{F_0-F'_0}\left[(2F'_0+1)(2F_0+1)\right]^{1/2}(2J+1)%
\;\left\{\begin{array}{ccc} S & I & F'_0\\ F' & 1 & J \end{array}\right\}%
\left\{\begin{array}{ccc} S & I & F_0\\ F & 1 & J \end{array}\right\}%
\label{3.16}%
\end{eqnarray}
\end{widetext}
and expressed by the sum of diagrams (\ref{c.11}) and (\ref{c.12}) in Appendix \ref{Appendix_C}. Such a cooperative in-scattering by both driving modes is constructed by various combinations of the amplitudes (a), (b) with the amplitude (c) in  Fig.~\ref{fig2}, and we distinguish this contribution from the optical pumping mechanism discussed above, and associate it with the CPT process. To activate this channel of incoherent scattering it is crucially important to precisely fulfill the conditions of two-photon resonance within a spectral resolution much narrower than $\gamma$ (i.e. for sufficiently long pulses, see preamble to Section \ref{Section_III}), such that the detunings of both the modes from the intermediate level are equal $\Delta_{nb}=-\Delta_{rn}$. Although the difference between dressed and undisturbed energies is not critical in the denominators of (\ref{3.14})-(\ref{3.16}), the exact resonance $\omega_1+\omega_2=\tilde{\omega}_{rb}$ is required in the defined transformation matrices.  

Similar terms, with an appropriate interchange of the indices, should be added to the master equation for the atom $B$ at the stage of its excitation. But again the situation is somewhat different and if the atom $A$ occupies the state $|r\rangle$ -- the CPT resonance condition cannot be created for the atom $B$ and in this case we have to eliminate the associated terms in its evolution. Nevertheless, there is a specific option when the atom $A$ is superposed between the states $|a\rangle$ and $|r\rangle$, and, as was pointed above, the density matrix has elements $\rho_{a,r;r,\beta}$ and $\rho_{r,\beta;a,r}$ with $\beta=b,a$. In this particular case we obtain  
\begin{eqnarray}
\dot{\rho}_{a,b;r,\beta}&=&\ldots -\frac{\gamma}{2}\sum_{n}\frac{\Omega_{rn}^{(2)\ast}\Omega_{nb}^{(1)\ast}}{4\Delta_{nb}^2}\,\rho_{a,r;r,\beta}(t)%
\nonumber\\%
\nonumber\\%
\dot{\rho}_{a,r;r,\beta}&=&\ldots -\frac{\gamma}{2}\sum_{n}\frac{\Omega_{rn}^{(2)}\Omega_{nb}^{(1)}}{4\Delta_{nb}^2}\;\phantom{\frac{}{}}\rho_{a,b;r,\beta}(t)%
\label{3.17}%
\end{eqnarray}
instead of (\ref{3.14}) and (\ref{3.15}), and 
\begin{widetext}
\begin{eqnarray}
\dot{\rho}_{a,m';r,m}(t)&=&\ldots+\gamma\,\exp\left[i\tilde{\omega}_{m'm}t\right]\,\sum_{\beta=a,b}\;%
\sum_{n',n}\exp\left[-i\tilde{\omega}_{b\beta}t\right]\frac{\Omega_{rn'}^{(2)\ast}\Omega_{n\beta}^{(1)\ast}}{4(-\Delta_{rn'})\Delta_{n\beta}}\,\rho_{a,r;r,\beta}(t)%
\sum_{q}C_{F'_0M'_0\, 1q}^{F'M'}C_{F_0M_0\, 1q}^{FM}%
\nonumber\\%
&&\times\;(-)^{F_0-F'_0}\left[(2F'_0+1)(2F_0+1)\right]^{1/2}(2J+1)%
\;\left\{\begin{array}{ccc} S & I & F'_0\\ F' & 1 & J \end{array}\right\}%
\left\{\begin{array}{ccc} S & I & F_0\\ F & 1 & J \end{array}\right\}%
\label{3.18}%
\end{eqnarray}
\end{widetext}
instead of (\ref{3.16}), which together with their Hermitian conjugated counterparts have to be added to the master equation describing the entire evolution.

The critical feature of the two-photon resonance process in the ladder-type system is its phase sensitivity. During the transient stage the combination of the stimulated and spontaneous coupling of $|r\rangle$ and $|b\rangle$ works towards rearrangement of the Rydberg coherence, under the steady state conditions that would turn the dynamics to irreversible conversion of the atomic subsystem to its dark state being a part of an eigenstate of the global system combining a two-mode field and an atom superposed between $|b\rangle$ and $|r\rangle$, which would further be insensitive to the excitation process. Conversion to the dark state in an ideal case of a closed three-level ladder configuration $|b\rangle\leftrightarrow|n\rangle\leftrightarrow|r\rangle$ would be most efficient if the spontaneous emission would transit the atom only via $|n\rangle$ to $|b\rangle$, as highlighted by diagram Fig.~2(d). In reality the situation is more subtle and there are several intermediate levels $n,n',\ldots$ forming several ladder-type transitions with different coupling strengths. Furthermore, in accordance with the protocol the created dark state is superposed with the state $|a\rangle$, and the trapping process competes with repopulation of the atoms out of the working channel by optical pumping. However if the driving modes are activated for both the atoms in a cw regime with infinite duration, a part of the atomic system would leak to the dark state consequently isolated from the interaction process. Eventually, with certain likelihood, that would contribute as a fraction of collective coherent state prepared for two atoms.\footnote{As was pointed out earlier the two-photon resonance and the effective optical pumping can be attained only asymptotically for much longer pulses than we consider here. But even the initial stage of both processes disentangles the qubits and can cause a considerable error in the quantum logic operations.} 

The difference of the CPT dark state formation process with the optical pumping phenomenon reveals that these mechanisms prevent the population imbalance differently and independently of each other. As can be straightforwardly verified by tracing the optical pumping terms for either atom $\mathrm{Tr}'_{A}(\dot{\hat{\rho}})_{\mathrm{OP}}=\mathrm{Tr}'_{B}(\dot{\hat{\rho}})_{\mathrm{OP}}=0$. Similarly for CPT time derivatives of both atoms we obtain: $\mathrm{Tr}'_{A}(\dot{\hat{\rho}})_{\mathrm{CPT}} = \mathrm{Tr}'_{B}(\dot{\hat{\rho}})_{\mathrm{CPT}}=0$.


\subsection{Other assumptions and the calculation scheme}
\noindent In addition to the decoherence processes discussed above there is a slow but unavoidable radiative decay channel of the Rydberg state itself. This process can be taken into consideration by including a small and empirically estimated exponential decay rate into equations (\ref{3.7}), (\ref{3.9}), (\ref{3.11}) and (\ref{3.12}). Then, due to the atom's cascade decay down to the ground state the process would eventually result in equal population of all the Zeeman sublevels belonging to the ground state. The realistic correction to the reduced density matrix can be constructed by an admixture of a respective Werner-type term proportional to a unit matrix compensating the depopulation of the working levels in such a decay process, with subsequent normalization.  

We conclude this section by describing our simulation algorithm. Initially both atoms are prepared in either $|a\rangle$ or $|b\rangle$ state, or a superposition these states. As a zeroth-order approximation the dynamical equations are solved for a particular linear momentum neglecting the inhomogeneity of the driving light beams. In this case, as shown in Appendix \ref{Appendix_B}, the transformation matrix (\ref{2.12}) can be found in an analytical form (\ref{b.2}). At the first-order approximation this result is corrected by keeping the differential terms in (\ref{2.8}) as perturbations via numerical solution of (\ref{2.8}). After these steps we have a realistically constructed set of probability amplitudes in (\ref{2.12}) at any time of the entangling process treated dynamically. As an undesirable variant at the end of the protocol one or even both the atoms can occupy the Rydberg state with small but non-vanishing probability. We have neglected these small elements of the density matrix in the analysis of incoherent losses but will further use them in our estimations of the correlation properties.

The dynamics of the reduced density matrix can be extracted by tracing the extended density matrix over the spatial variables. The tricky point is that the trace has to be evaluated in the basis of harmonic oscillator eigenfunctions and over the Gibbs measure at a given temperature. This part of the calculation can be done numerically only for low temperatures when the thermal state occupies a few low energy oscillator modes. This is however the most desirable limit for potential applications, so we do not attempt to push the calculation too far in the higher-temperature regime. Eventually at this stage we have recovered the reduced density matrix as a function of time during the entanglement protocol from its beginning up to its end. That gives us the starting point for further inclusion of the incoherent losses.  

The incoherent losses can be realistically estimated by straightforward numerical evaluation of the increments for the density matrix, accumulated during the process. The increments are small but expressed by finite integrals of the respective time derivatives described in the preceding section. We substitute the density matrix, as approximately reproduced by dynamical solution, in these integrals. These corrections do not violate the normalization condition for the density matrix. At the final step we correct the result by incorporating the direct decay process of the Rydberg state as described above. 

\section{Results} \label{Section_IV}

\noindent In this section we present the results of our numerical simulations for the basic benchmarks of the entanglement protocol, namely, for purity and fidelity of the prepared entangled state of two qubits and for the truth table of the CNOT gate implemented with the protocol. We analyze the results and compare the cases of different experimentally accessible excitation geometries searching for an experimental configuration optimizing the entanglement preparation.

\subsection{The excitation geometries}

\noindent To realize the protocol experimentally, it is necessary to fulfill the conditions of a closed two-level transition between the qubit state $|b\rangle$ and the Rydberg state $|r\rangle$. This guarantees that the required $\pi$- and $2\pi$-pulses between the ground and excited states may be realized exactly. Otherwise the transitions between $|b\rangle$ and several $|r\rangle$-states would repopulate the atoms out of the main channel and involve more states in the interaction process. Since the hyperfine structure of the Rydberg states is unresolved within the protocol duration, typically within $1\,\mu s$, it is crucially important to select the excitation to a single Zeeman state $|r\rangle$, which can be further specified in the basis of the total electronic and nuclear spin angular momenta. 

\begin{figure}[tp]
\includegraphics[width=8.6cm]{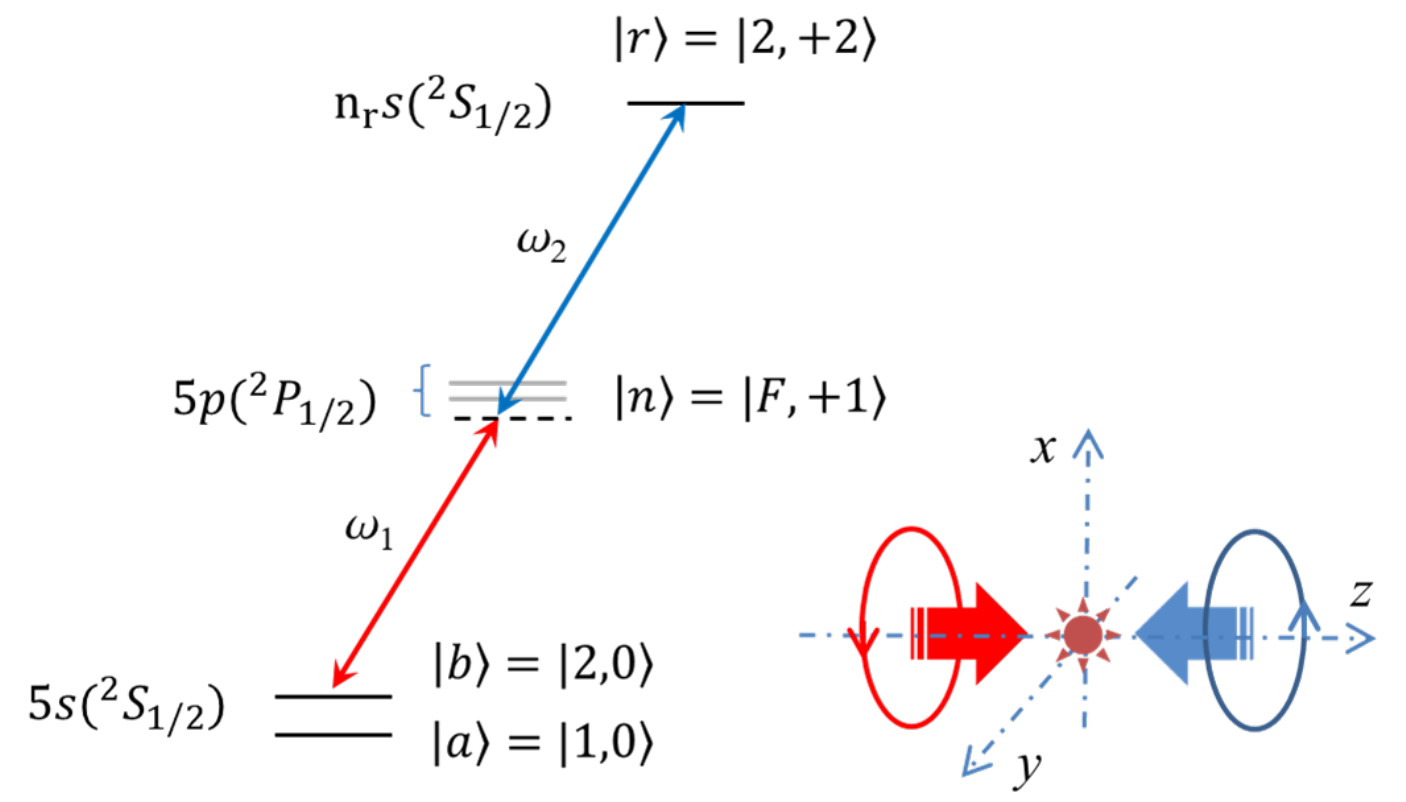}
\caption{The transition diagram and excitation geometry for ${}^{87}$Rb driven by two counter-propagating and circularly polarized light beams. In the diagram the participating states are specified by the concrete numbers of the total electronic and nuclear spin angular momenta $F_0=2,\,F=1,2,\, F_{r}=2$ and their projections $M_0=0,\,M=+1,\,M_{r}=+2$. The used energy configuration of ${}^{87}$Rb is effectively two-level and provides coupling of the qubit state $|b\rangle$ only with a single Rydberg state $|r\rangle$ having principal quantum number $\mathrm{n}_r \simeq 50-100$.}
\label{fig3}%
\end{figure}

As a first example we consider the specific excitation channel existing only in the energy manifold of ${}^{87}$Rb when two circularly-polarized beams provide the coupling between the ground and Rydberg electronic states, both having zero orbital momenta, as shown in Fig.~\ref{fig3}. We associate the axial direction of the trap with the $z$-axis and assume atoms $A$ and $B$ as located in the $x,y$-plane, and define the respective directors  $\mathbf{e}_x,\,\mathbf{e}_y,\,\mathbf{e}_z$. The qubit is encoded into the hyperfine clock transition and we can specify the qubit states precisely as $|a\rangle = |5s(^2S_{1/2});F_0=1;M_0=0\rangle$ (logic ``$0$'') and $|b\rangle = |5s(^2S_{1/2});F_0=2;M_0=0\rangle$ (logic ``$1$''). 

Then the qubit state $|b\rangle$ can be coupled by a two-photon transition with the upper state $|r\rangle = |\mathrm{n}_{r}s(^2S_{1/2}); F_r = 2; M_r = +2\rangle$ with the principle quantum number $\mathrm{n}_{r}\sim 50-100$. This state plays a role of quantum bus blocking the double excitation of two atoms. We do not fix a concrete value of $\mathrm{n}_r$ since in our simulations it contributes to the energy shift $\hbar\delta_{R}$ only, which we consider as an external and independent parameter. The choice of the Rydberg states with zero orbital angular momentum is additionally motivated by a convenient isotropic structure of this shift. Excitation with two laser beams, both having the same circular polarizations (but different helicities) $\mathbf{e}_1 = \mathbf{e}_2 = \mathbf{e}_{+1}=-(\mathbf{e}_x+i\mathbf{e}_y))/\sqrt{2}$, couples the selected ground and Rydberg Zeeman states $|b\rangle$ and $|r\rangle$ via two intermediate states $|n\rangle = |5p(^2P_{1/2}); F = 1,2; M = +1\rangle$. 

\begin{figure}[tp]
\includegraphics[width=8.6cm]{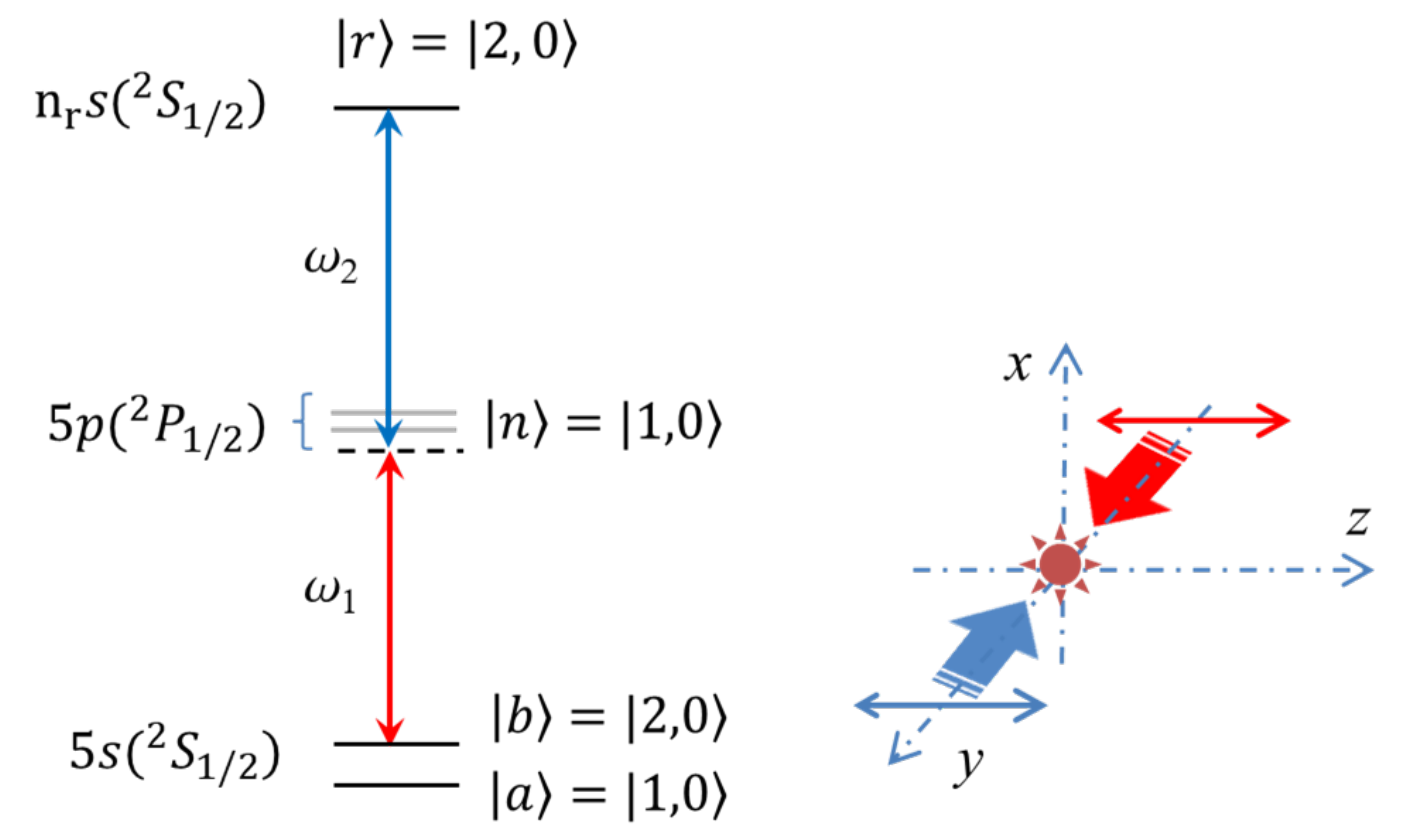}
\caption{Same as in Fig. \ref{fig3} but for the excitation by two linearly polarized light beams propagating in the transverse direction. The energy configuration is also effectively two-level due to specific selection rules for electric dipole transitions.}
\label{fig4}%
\end{figure}

Another example of the excitation process, which we shall consider, is shown in Fig.~\ref{fig4} where the two counter-propagating driving beams are directed along $y$-axis and linearly polarized along the $z$-axis. In this case the recoil linear momentum pushes each of the atoms in the transverse plane where they have tighter confinement than in the axial direction. The transverse motion can be frozen with the aid of the Raman sideband cooling (RSBC) protocol and the negative influence of the recoil on the entanglement preparation, explained in Section \ref{Section_IID}, can be minimized. Let us point out here that RSBC down to the trap ground state in a three-dimension configuration is quite challenging and it would be problematic to do it in an atomic lattice consisting of many qubits, see \cite{PorozovaPRA2019}. There is no need to cool the axial motion for the excitation configured with the linearly polarized light beams in geometry of Fig.~\ref{fig4}, as we will explain later.

Unlike the excitation channel shown in Fig.~\ref{fig3} the excitation by linearly polarized light beams with $\mathbf{e}_1 = \mathbf{e}_2 = \mathbf{e}_{z}$ can be used for any alkali-metal atom. An important advantage of the linear polarizations is in convenient selection rules for the dipole coupling of the Zeeman states with a zero projection of the angular momentum, which provides that only one intermediate state $|n\rangle = |5p(^2P_{1/2}); F = 1; M = 0\rangle$ contributes to the ladder-type two-photon excitation. 

\subsection{Fidelity of entanglement and purity of the prepared state}

\noindent The two-particle density matrix of atoms $A$ and $B$ was calculated for geometries of Figs.~\ref{fig3} and \ref{fig4} as described in Sections \ref{Section_II} and \ref{Section_III}. For the open system, when interaction with the environment only slightly disturbs its dynamical behavior, the deviation from the ideal state (\ref{2.17}) can be expressed by fidelity of this state and the actually prepared mixed state of two atoms ${\cal F} = \langle \psi|{\hat \rho}|\psi \rangle_{AB}$. The mixed state $\hat{\rho}$ can have an eigenfunction $|\psi \rangle$ with a maximal eigenvalue, which can be different from $|\psi \rangle_{AB}$. In this case we will use the purity ${\cal P} = {\rm Sp} \hat{\rho}^2$ as an intrinsic parameter indicating the priority of the dynamical behavior in the state preparation.     

The subtle point is that for correct comparison based on a fidelity criterion we should eliminate the extra phases associated with the light shifts induced by the driving lasers to the hyperfine sublevels in the prepared state. Indeed the amplitudes of both the qubit states $|a\rangle$ and $|b\rangle$ of both the atoms accumulate the phases during the protocol, see (\ref{2.6}) and related comments. For the calculated fidelity ${\cal F}$ we have eliminated these extra phase shifts, which in an experiment would be compensated for by additional spin rotations realized with additional single qubit rotations.

The entanglement protocol, described in Section \ref{Section_IIC}, is divided into three subsequent transformations by $\pi$, $2\pi$ and $\pi$ pulses. We express the duration of the $\pi$-pulse via an effective Rabi frequency as $\tau_\pi = \pi/|\Omega|$, and similarly for a $2\pi$ pulse with equal $\Omega$'s for all the pulses, such that the full protocol duration is given by $2\tau_\pi+\tau_{2\pi}$. Then, the longer duration corresponds to the smaller effective Rabi frequency. Note that our calculations are sensitive to both Rabi frequencies of the driving beams $\Omega^{(1)}$ and $\Omega^{(2)}$, which may be varied independently. Here we use $\Omega^{(1)} \simeq \Omega^{(2)}$ but certain further optimization of the mutual relation between these quantities is possible.

\begin{figure}[tp]
\includegraphics[width=8.6cm]{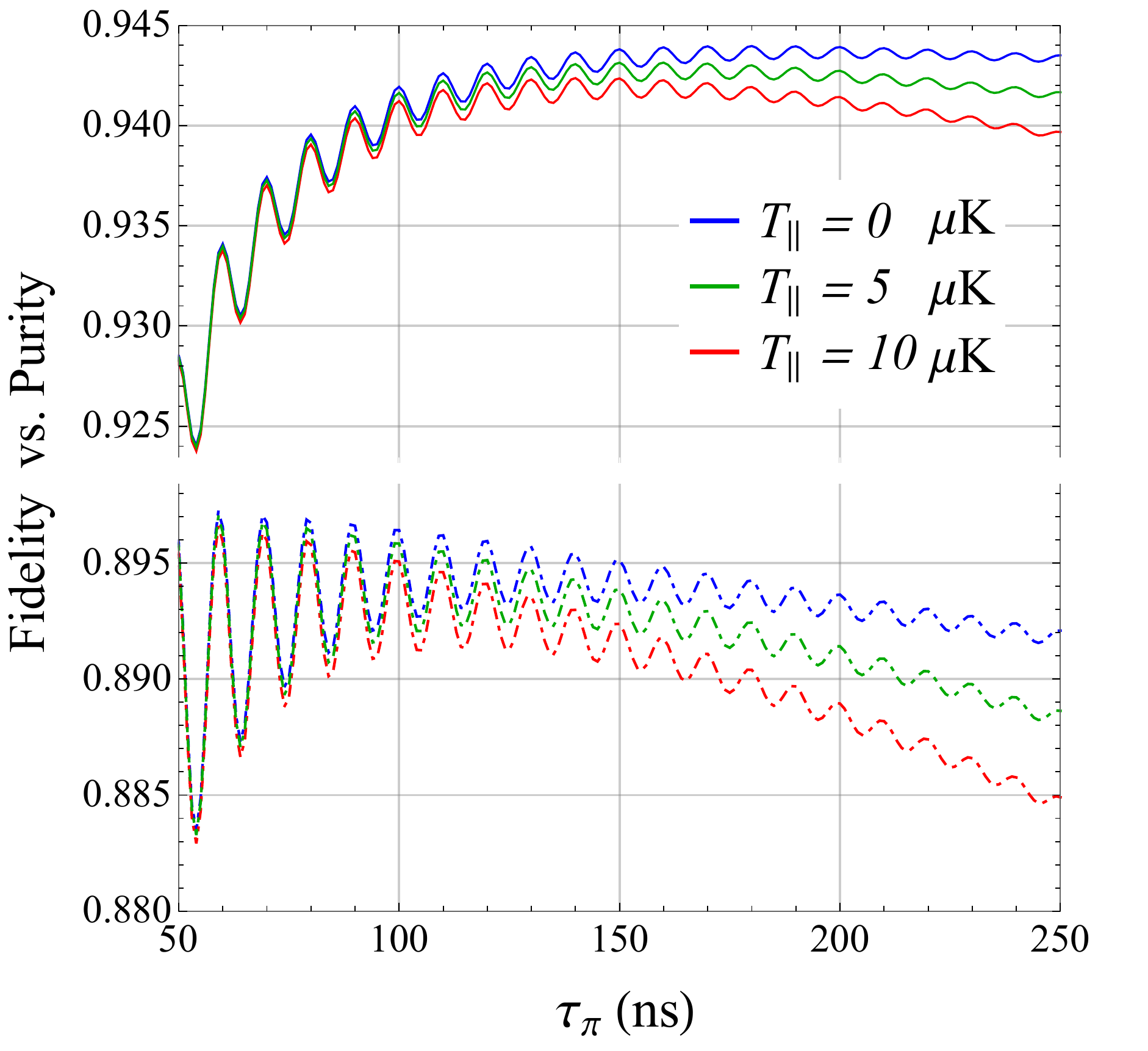}
\caption{Fidelity ${\cal F}$ (solid curves) and purity ${\cal P}$ (dashed curves) of the entangled state, prepared by the protocol of Rydberg blockade with excitation by circularly polarized light beams, see Fig.~\ref{fig3}, and considered as a function of the $\pi$-pulse duration $\tau_\pi = \pi/|\Omega|$. Time beats in these dependencies indicate a small but non-negligible probability of an off-resonant transition and simultaneous occupation of $|r\rangle$ states by atoms $A$ and $B$. The transverse degrees of freedom are frozen for both of the atoms, but the axial mode is thermalized with a variable temperature $T_\parallel = 0,\,5,\,10$ $\mu K$.}
\label{fig5}%
\end{figure}

\begin{figure}[tp]
\includegraphics[width=8.2cm]{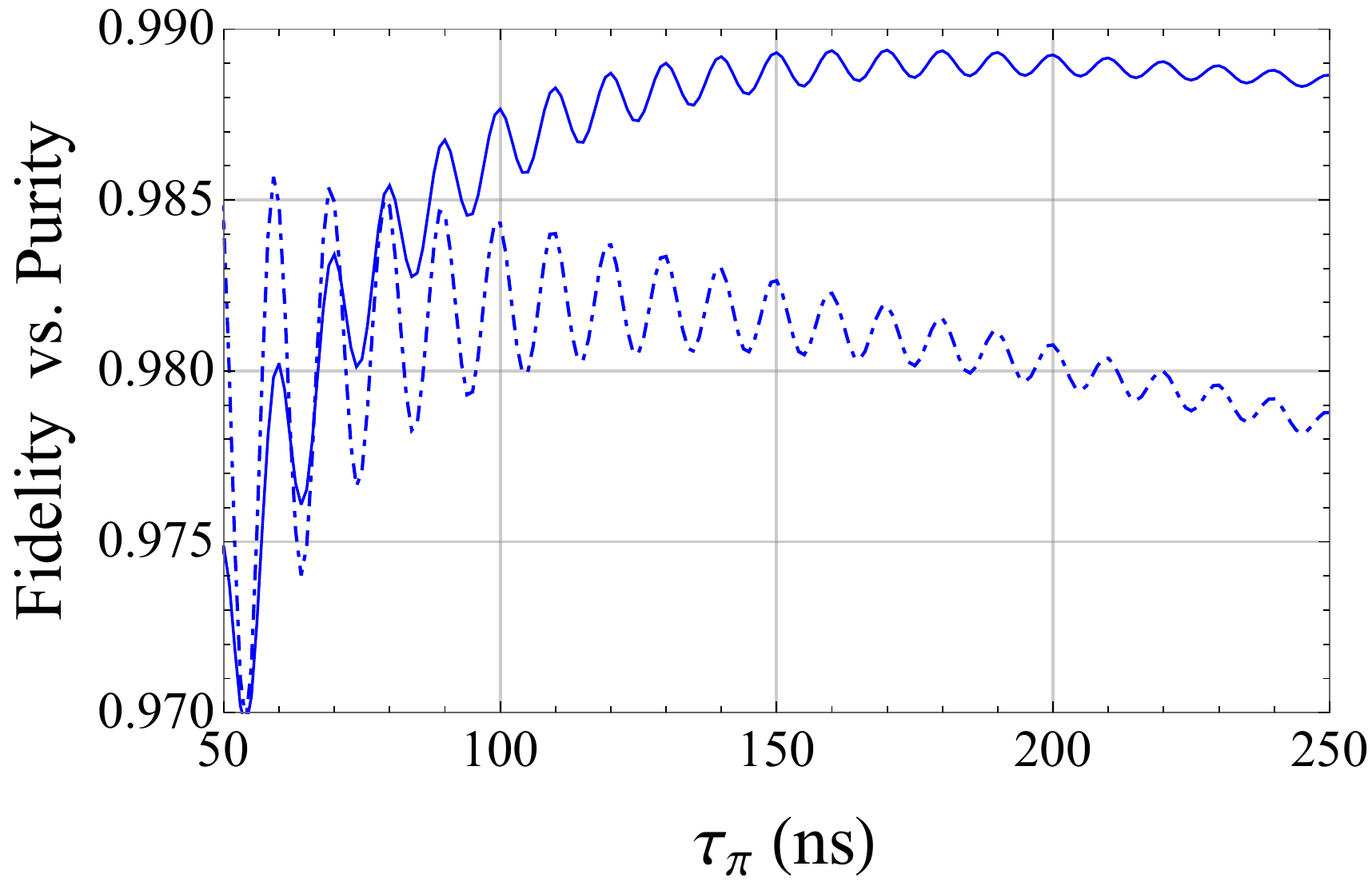}
\caption{Same as in Fig. \ref{fig5} but for the excitation geometry of Fig.~\ref{fig4}. The dependencies, related to different temperatures $T_\parallel = 0,\,5,\,10\, \mu K$, are unresolved within the plot scale.}
\label{fig6}%
\end{figure}

The parameters ${\cal F}$ and ${\cal P}$, calculated as a function of $\tau_\pi$, are shown in Figs.~\ref{fig5} and \ref{fig6} for the excitation geometries shown in Figs.~\ref{fig3} and \ref{fig4}, respectively. The counter-propagating $780$~nm and $480$~nm light beams are detuned by $\Delta_{nb} = - 2\pi \cdot 3000$~MHz from the intermediate state $|n\rangle = |5p(^2P_{1/2}); F = 1;M\rangle$. \footnote{Other parameters of the dipole trap, used in our calculations, are the same as in \cite{SingleQubit2021}.} We assume that both of the atoms have their radial degrees of freedom cooled and occupy the ground state of the transverse vibrational modes. The axial mode exists in a thermal state described by the Gibbs measure with a temperature varied as $T_\parallel = 0$, $5$ and $10$ $\mu$K. The temperature dependence is mainly observed in Fig.~\ref{fig5} for the pulses of longer duration and is unresolved in Fig.~\ref{fig6} within the graph scale for the tested calculation domain. 

The quantities ${\cal F}$ and ${\cal P}$ demonstrate quite an intriguing behavior as functions of the protocol duration for different excitation conditions. There are time beats of ${\cal F}$ and ${\cal P}$ vanishing for a longer duration. That is a consequence of the protocol imperfection allowing the simultaneous occupation of the Rydberg state by atoms $A$ and $B$, known as blockade leakage. As commented in Appendix \ref{Appendix_B} for such an event the respective probability is small and oscillates with the frequency, given by (\ref{b.5}) and roughly estimated as $2\delta_R$. This is visualized on the graphs of Figs.~\ref{fig5} and \ref{fig6}, and, in accordance with (\ref{b.2}), the transition amplitude (reproduced by the amplitude of time beats in the graphs) vanishes for smaller values of the effective $|\Omega|$ i.e. for the longer protocol duration. 

In our calculations we set the frequency shift $\delta_R$ of the energy levels for a doubly excited Rydberg state, as $\delta_R = 2\pi\,\cdot 50\,\mathrm{MHz}$, which realistically estimates the dipole interaction for a pair of rubidium atoms, excited to the upper state with a principal quantum number $\mathrm{n}_r\sim 50-100$, and separated by a distance of several microns, see \cite{SaffmanWalkerMolmer2010}. Although the condition $|\Omega|=\pi/\tau_\pi\ll\delta_R$ is fulfilled but, as follows from our calculations, the probability amplitude of double excitation to the Rydberg state is not negligibly small and for a shorter protocol duration it increases, which eventually results in leakage of the system out of the controllable dynamics and induces an extra phase shift between the basis states. Thus the dependencies of Figs.~\ref{fig5} and \ref{fig6} suggest the optimum for the effective Rabi frequency and the protocol duration near $\tau_{\pi}\sim 100-200$ ns for both the considered geometries.

Furthermore to clarify the mutual relation between ${\cal F}$ and ${\cal P}$ let us approximate the density operator by the following Werner-type mixture
\begin{equation}
    {\hat \rho} \sim (1-x)|\psi\rangle\langle\psi| + \frac{x}{4} \hat{I}_{AB},
    \label{4.1}%
\end{equation}
where we associate the state $|\psi\rangle$ with the eigenstate of $\hat{\rho}$ corresponding to the largest eigenvalue and then admix it with a maximally mixed state described by a unit matrix $\hat{I}_{AB}$ in the linear span of states $|a\rangle$ and $|b\rangle$ for two atoms. In the considered case of a small $x$ we can expect that $|\psi\rangle\sim|\psi\rangle_{AB}$. If these functions coincide, we straightforwardly obtain that ${\cal F}\geq{\cal P}$ and equality is only possible for $x=0$. That is perfectly confirmed by the dependencies plotted in Fig.~\ref{fig5}. But for a rather short protocol duration we arrive at the opposite inequality ${\cal F}<{\cal P}$, as is clearly visible in Fig.~\ref{fig6}. This unambiguously tells us that the excitation by linearly polarized light beams provides a nearly dynamical behavior, with ${\cal P}\to 1$, but at the same time indicates a deviation between the prepared state $|\psi\rangle$ and the ideal state $|\psi\rangle_{AB}$. This difference can be quite important for further implementation of quantum logic operations. In general we always have $\langle\psi_{AB}|\psi\rangle \neq 1$ since even under the dynamical evolution the atoms can leak out of the main channel of Rydberg blockade, and state $|\psi\rangle$ will contain $|\psi\rangle_{AB}$ only as a part of its Schmidt decomposition. For optimization of the logical operation it would be wise to avoid the poorly controllable domain with relatively high $|\Omega|\lesssim\delta_R$ and short duration $\tau_{\pi}\gtrsim\delta^{-1}_{R}$.

In the opposite limit of infinitely long pulses, the processes of incoherent losses, discussed in Section \ref{Section_III}, irreversibly damage the generated entanglement. Asymptotically, the dependencies of Figs.~\ref{fig5} and \ref{fig6} should approach the limits of the completely randomized mixed states, i.e. ${\cal F}, {\cal P}\to 1/g_{AB}$, where $g_{AB}$ is the degeneracy of the ground state in the combined system of two atoms. In our numerical simulations this asymptotic behavior cannot be verified due to the restrictions of the considered model. But it seems more important that in the optimal region with $\tau_{\pi}\sim 100-200$ ns, fairly reproduced by our model, there is a significant difference in estimates of parameters ${\cal F}$ and ${\cal P}$ for the excitation geometries, designed either with circular polarizations (Fig.~\ref{fig3}), or with linear polarizations (Fig.~\ref{fig4}).   

As was mentioned in Section \ref{Section_IID}, entanglement is reduced for the two photon excitation acting within a finite time interval due to the recoil of linear momentum, which mainly affects the control atom $A$ spending a relatively long time $\tau_{2\pi}$ in its Rydberg state. Referring to our estimate (\ref{2.20}) above, we can point out that roughly the displacement $\boldsymbol{\alpha}\propto(\hbar \mathbf{q} +\mathbf{p})/m\;\tau_{2\pi}$ being averaged over thermal distribution for momentum $\mathbf{p}$; it is visualized as a path segment directed along the transferred linear momentum $\hbar \mathbf{q}$.\footnote{Note that in our estimates the recoil linear momentum is only negligibly varied with the principle quantum number of the Rydberg state with $\mathrm{n}_r\gg 1$.} The direction is different for  Figs.~\ref{fig3} and \ref{fig4} but the segment length is approximately the same, such that the recoil affects the motion in the transverse and axial modes of the trap oscillator more or less similarly. That is predicted by the observed slight dependence of the process on the thermal state of the axial mode. Note that for the trap oscillator the dimensionless displacement $\boldsymbol{\alpha}$, scaled by the position uncertainty, is even higher for the tightly confined transverse mode than for the loose axial mode. But in spite of this, the quality of entanglement is certainly better for the dependencies of Fig.~\ref{fig6} than Fig.~\ref{fig5}. So the recoil plays no dominant role in reduction of the entanglement under the conditions considered here. The recoil effects will become a major error source if the incoherent scattering from the intermediate state is significantly reduced \cite{Saffman_PRA2021}. 

The optimal duration with $\tau_\pi \sim 100 - 200$ ns is mainly provided by a trade off between the losses coming from incoherent scattering, which are increased with the increased duration of the interaction process, and uncontrollable deviations from the target state $|\psi\rangle_{AB}$ appearing for short control pulses. The incoherent losses reduce the fidelity and purity differently for the considered excitation geometries. The optimal attained fidelity of the prepared entangled state is varied from $\sim 95\%$ in the case of Fig.~\ref{fig3} to $\sim99\%$ in the case of Fig.~\ref{fig4} and that highlights the advantage of two photon excitation with linear polarizations. The excitation by linearly polarized light beams provides convenient selection rules with only one intermediate state involved in the two-photon interaction process. That minimizes the negative contributions of the processes discussed in Section \ref{Section_III}.

The numerical results presented in this section were primarily focused on reproducing our experimental limitations and it eventually bounded the fidelity of the entanglement protocol by $99\%$ at highest. However this technical benchmark may be improved by designing the two-photon excitation via higher intermediate states, for example via $6p(^2P_{1/2})$ in the case of rubidium atoms. This state has a weaker spontaneous decay than $5p(^2P_{1/2})$ and the incoherent losses might be smaller. Such excitation channels were realized in experiments \cite{Lukin_PRL2019,Saffman_PRL2019}. The negative effect of incoherent scattering can be also suppressed by increasing the Rydberg beams detuning from the intermediate state at fixed effective Rabi frequency $\Omega$. 

\subsection{The truth table for a CNOT gate}


In the most general case, a three-qubit quantum logic operator is required to construct an arbitrary quantum network that includes all of the options of classical computations \cite{Deutsch1989}. However for the widely used and universal set of quantum computations, proposed in \cite{Weinfurter1995}, the data processing can be realized by compilation of the two-qubit CNOT gates with arbitrary single-qubit rotations. Then any unitary transformation realized by a quantum computer, can be expanded as a finite set of subsequent transformations involving only the CNOT and single-qubit operations. Although the CNOT and CZ gates may both be used in such universal gate sets, the CNOT is operationally preferable, since its truth table may be directly observed in the computational basis without full gate tomography. Unfortunately, given various imperfections discussed above the CNOT gate cannot be implemented ideally and in this section we describe how the above discussed decoherence processes affect the quality of the CNOT gate, applying our calculations to the output density matrices. 

The transformation between CZ and CNOT gates is physically implemented by two additional microwave pulses providing the single-qubit $\pi/2$-rotations on the Bloch sphere of the target atomic qubit $B$. Since fidelity of the microwave single-qubit gates is typically much higher than fidelity of the entangling Rydberg gate \cite{Zhan_PRL2018}, we simulate them as an infinitely short lossless transformations in the linear span of the computational basis $|aa\rangle$, $|ab\rangle$, $|ba\rangle$, $|bb\rangle$ ignoring the option for the state decoherence during such an infinitely short single-qubit operation. The complete transformation sequence includes the Rydberg blockade realization of the CZ gate with many channels of losses and therefore takes part of the system off the computational subspace. In those situations when either one or both the atoms leave this subspace, we trace the density matrix over the spin states of the lost atom[s]. Then we can define the following conditional probabilities clarifying the figure of merit of the quantum gate
\begin{eqnarray}
P(|\alpha,\beta\rangle)&=&\rho_{\alpha,\beta;\alpha,\beta}\ \ \ \ \ \ \ \ \mathrm{if}\ \ \alpha,\beta\in (a,b)%
\nonumber\\%
\nonumber\\%
P(|\alpha,\varnothing\rangle)&=&\sum_{\beta\neq a,b}\rho_{\alpha,\beta;\alpha,\beta}\ \ \mathrm{if}\ \ \alpha\in (a,b)%
\nonumber\\%
P(|\varnothing,\beta\rangle)&=&\sum_{\alpha\neq a,b}\rho_{\alpha,\beta;\alpha,\beta}\ \ \mathrm{if}\ \ \beta\in (a,b)%
\nonumber\\%
P(|\varnothing,\varnothing\rangle)&=&\sum_{\alpha\neq a,b}\sum_{\beta\neq a,b}\rho_{\alpha,\beta;\alpha,\beta}%
\label{4.2}%
\end{eqnarray}
where we have denoted the absence of the particular atom in the computational subspace by the symbol of an empty set formally written in Dirac notation in the function argument, expressing the respective possibility. The leakage from this subspace results either from the repopulation process due to the incoherent scattering of the driving modes on any atom, or from spontaneous emission from the Rydberg state of atom $A$. The density matrix is calculated for the varied initial conditions when each atom subsequently occupies the particular computational state either $|a\rangle$ or $|b\rangle$.  

\begin{figure}[tp]
\includegraphics[width=8.6cm]{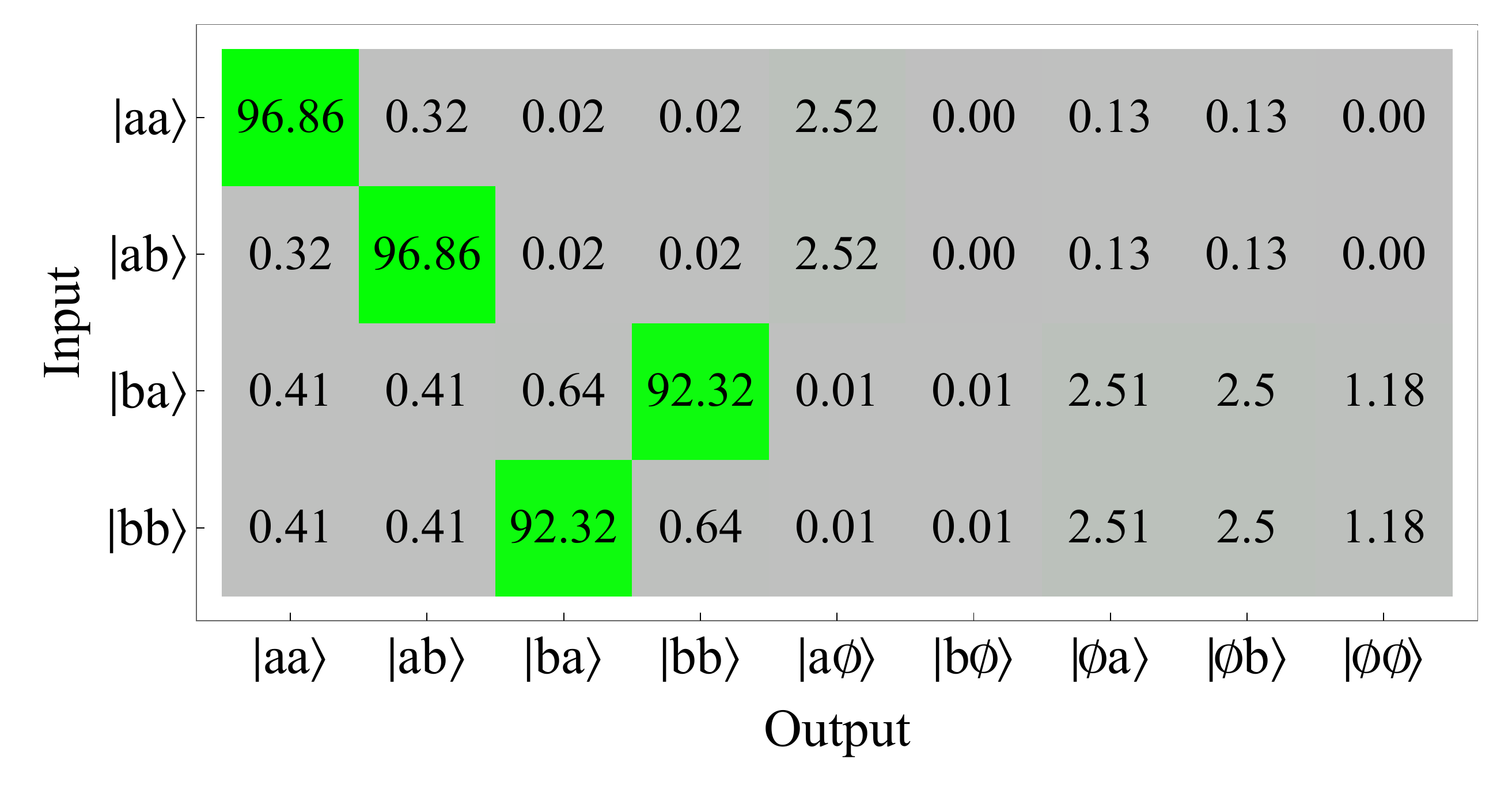}
\caption{The truth table for the CNOT quantum gate calculated for the $\pi$-pulse duration $\tau_\pi \simeq 150 ns$ and for the excitation geometry shown in Fig.~\ref{fig3}. The axial mode is thermalized with the temperature $T_\parallel = 10$ $\mu$K and other parameters are the same as in Fig.~\ref{fig5}. The conditional probabilities are defined by (\ref{4.2}) and shown in ($\%$). The table rows specify the input states and the table column the outputs, see the text for more details.}
\label{fig7}%
\end{figure}

\begin{figure}[tp]
\includegraphics[width=8.6cm]{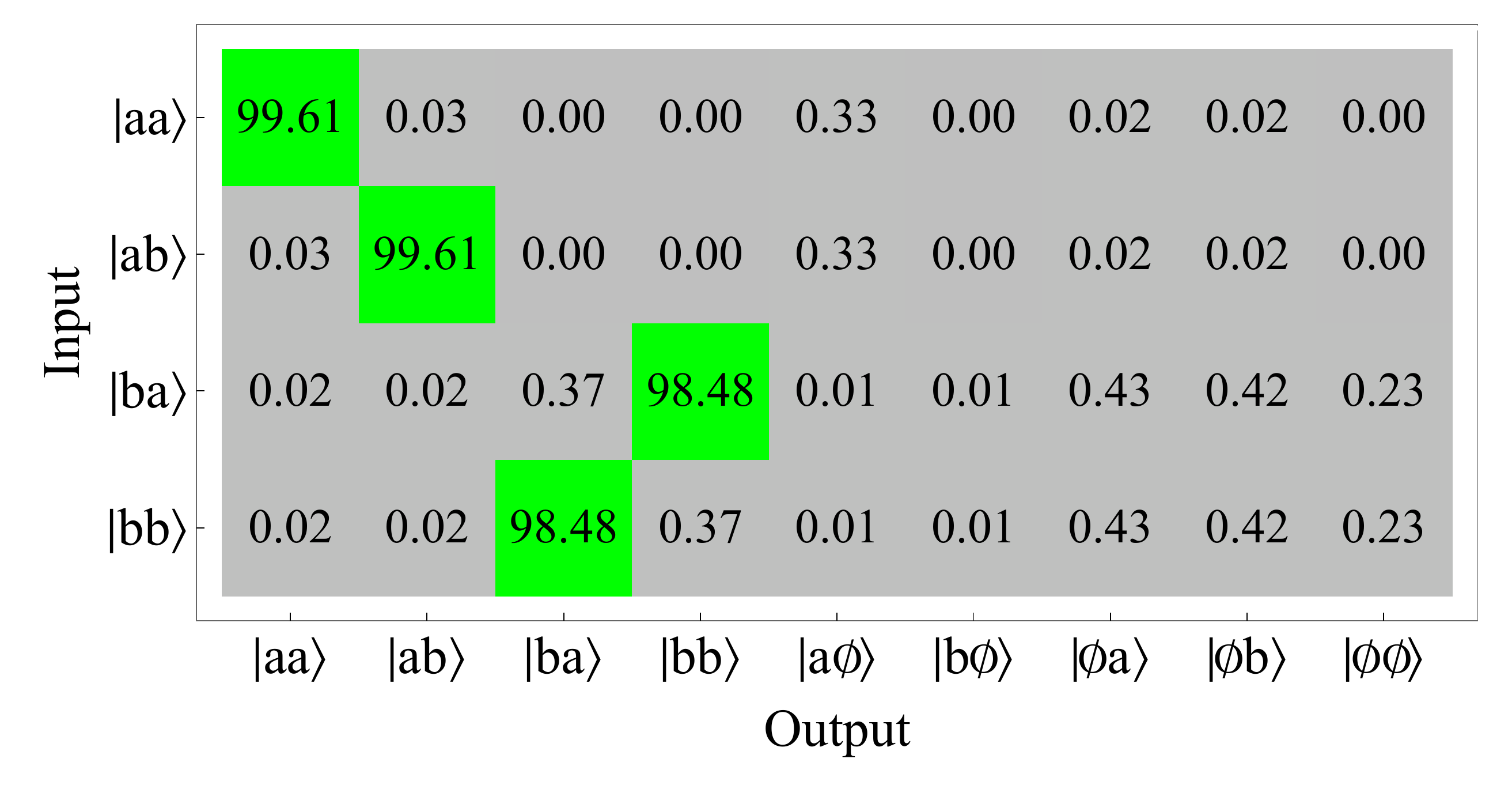}
\caption{Same as in Fig.\ref{fig7} but for the excitation geometry shown in Fig.~\ref{fig4}.}
\label{fig8}%
\end{figure}

The above defined conditional probabilities estimate the likelihoods of different ``outputs'' given different originally prepared ``input'' states, and they can be organized in a matrix known as the CNOT truth table. In Figs. \ref{fig7} and \ref{fig8} we present the truth tables for the excitation geometries shown in Figs. \ref{fig3} and \ref{fig4} respectively. The numbers in the table cells reproduce the probabilities (\ref{4.2}) where the output possibilities and the different initial conditions are specified by the table column and rows, respectively. 

The numerical results presented in Fig.~\ref{fig8} show that the excitation by linearly polarized driving modes (Fig.~\ref{fig4}) provides a certain advantage for the CNOT logic operation. As pointed out earlier, the reason for this is in specific selection rules for electric dipole transitions that prevent an undesirable incoherent repopulation (optical pumping) of the atoms between the computational states. The spontaneous Raman transitions from $|a\rangle$ to $|b\rangle$ and $|b\rangle$ to $|a\rangle$ are forbidden and do not affect the data processing. That would not be the case for the geometry of Fig.~\ref{fig3}. However, for both geometries there are additional small but non-negligible transition probabilities from the collective states $|ba\rangle$, $|bb\rangle$ to $|aa\rangle$ and $|ab\rangle$ respectively, since the signal atom $A$ can spontaneously decay during the protocol from its Rydberg state to any Zeeman sublevel of its ground state with the lifetime of $\sim 100$ $\mu$s.

\begin{figure}[tp]
\includegraphics[width=8.6cm]{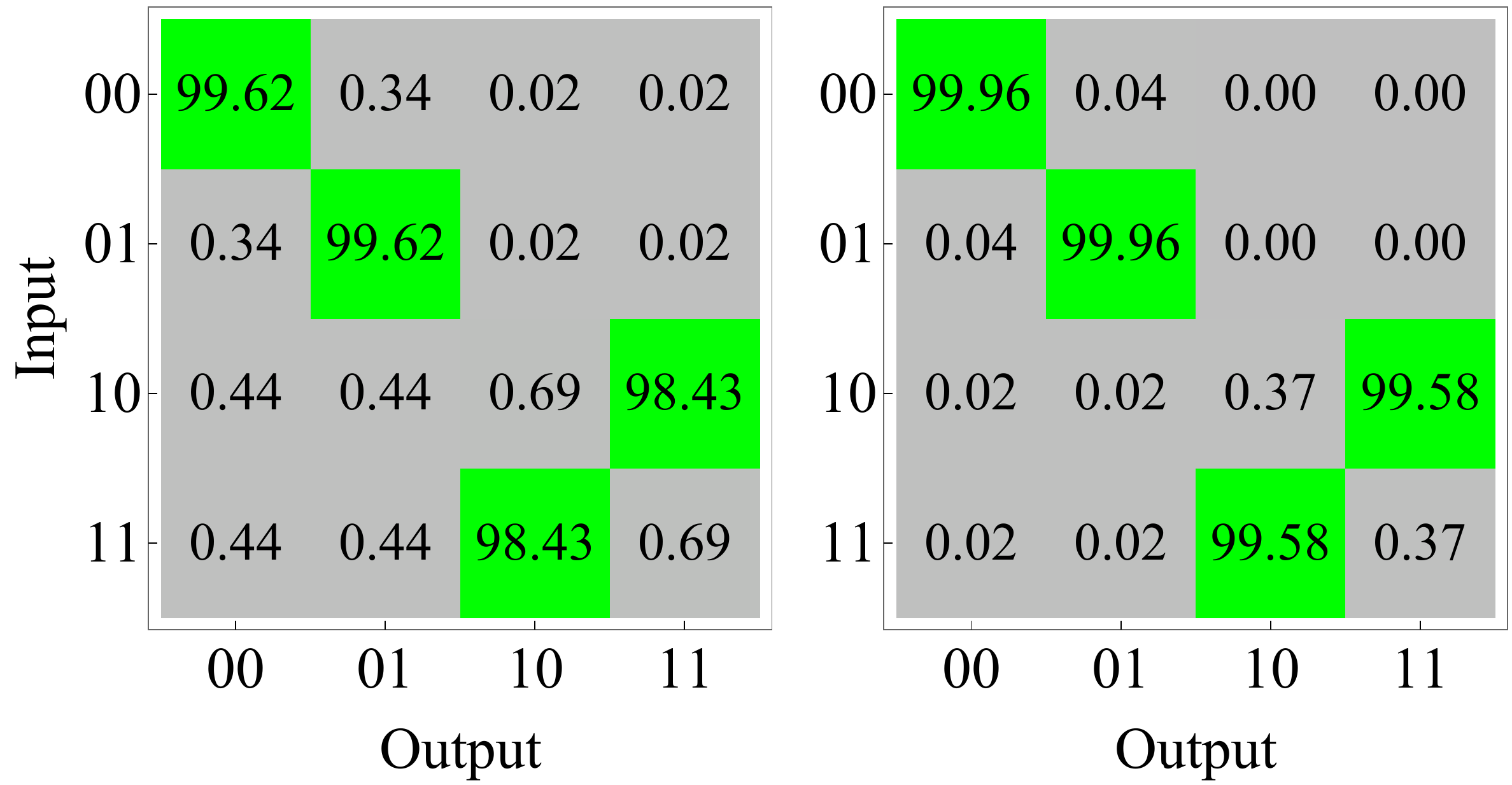}
\caption{The truth tables for the CNOT based quantum gate, calculated for the same parameters as in Figs.~\ref{fig7}, \ref{fig8}, but after post-selection of the atoms in the computational subspace. The tables are shown for both considered excitation geometries: Fig.~\ref{fig3} (left) and Fig.~\ref{fig4} (right).}
\label{fig9}%
\end{figure}

The truth tables can be corrected by additional post-selection verifying the existence of both the atoms in the computational subspace. We skip the discussion of how that could be implemented technically and associate the correction with the renormalized projection of the density matrix onto the computational subspace.  We present the CNOT truth tables illustrating the post-selected data processing in Fig.~\ref{fig9}. The tables express the quantum gate operation directly with the numbers $(00)$, $(01)$, $(10)$, $(11)$ and give us upper estimates of the figure of merit for the considered realization of the CNOT gate. Further optimizations are surely possible but significant improvement of the protocol parameters towards gate errors significantly less than $0.1\%$ would challenge us to search for other physical solutions.

A complete description of the quantum gate in the logical subspace is given by a process matrix which may be reconstructed from our numerical simulations, see appendix \ref{Appendix_D} for details.

Let us draw attention to the fact that the excitation geometry shown in Fig.~\ref{fig4} revealing certainly better characteristics than the one in Fig.~\ref{fig3} has not been experimentally verified so far. In experiment it would be not so easy to do since, in accordance with the protocol, the multi-qubit quantum register, structured in the transverse plane, should be provided with the possibility of individual addressing for each qubit. That favors experimental configurations which naturally imply the driving beams directed along the axial axis, i.e. orthogonal to this plane, with tight focusing of the beams with waists of a few microns. Nevertheless, some alternative solutions for selective addressing were implemented experimentally in \cite{Browaeys2016,Whaley2004}, which are suitable for the excitation geometry suggested by Fig.~\ref{fig4}.

We conclude this section with the following remark. Observation of collective dynamics of the atoms driven by a Hamiltonian with tunable and controllable parameters is an essential element of a quantum simulator utilized for studying many-body physics, phase transitions, quantum chemistry, etc., as well as for developing universal quantum computation. 
As follows from our simulations presented for the system of two atomic qubits, the combination of optimal excitation geometry with the Raman side-band cooling of the atoms' transverse spatial motion only may significantly improve the performance of the entanglement protocol based on Rydberg blockade. There are justified expectations that this key result could be extended on a multi-qubit system and be applicable for any alkali-metal atoms and finally represent an effective tool for the creation of large-scale entanglement in atomic systems.


\section{Conclusion}
\noindent We have analyzed various physical mechanisms underlying the protocol of the atomic spin entanglement by the Rydberg blockade technique and verified the proposed model by numerical simulations. Unlike many previous studies, mostly focusing on quantum simulators operating with multi-qubit systems, here we were motivated by clarifying the main physical barriers in attaining an ideal scenario for digital quantum data processing. Although fidelity for a two-qubit entanglement at the level better than 95\% was reported in some advanced experimental demonstrations, see \cite{Lukin_PRL2019, Saffman_PRL2019}, we have obtained and discussed many difficulties for its further improvement within technically limited capabilities of a currently used experimental design. 

The main source of errors lies in spontaneous scattering which unavoidably follows coherent dynamical coupling to the Rydberg states utilized to realize quantum logical operations. Our numerical simulations, based on realistic description of the entire interaction process, suggest optimal duration of the excitation pulses and control field amplitudes, which minimize the negative influence of various channels of incoherent scattering. As an important technical option for practical optimization of the protocol, we have demonstrated the advantages of using linear $\pi$-polarized excitation beams. That would eliminate part of the spontaneous loss channels, and, as verified by our numerical simulations, would improve the basic parameters such as fidelity and purity of the prepared entangled states. Interestingly, under the experimental conditions considered here, the recoil effect plays no dominant role in the reduction of gate fidelity. However, it will ultimately limit the achievable gate fidelity at higher values, when other imperfections such as incoherent scattering from the intermediate state are eliminated \cite{Saffman_PRA2021}. 


Here we focused on the original variant of the blockade gate, while current realizations tend to modify the protocol both for technical reasons and to reduce the error rate, however the main sources of errors and the physical model behind our analysis remain the same, so it can be easily modified for other blockade-type quantum gates. We leave the relative analysis and comparison of performance of other blockade-based gates for future work. Focusing on fundamental limitations we have also not accounted for the influence of technical noise, such as phase and amplitude noise of the excitation lasers, fluctuating electrical fields, etc., these noise sources may be incorporated in the model later as random fluctuations of the classical control parameters \cite{Lahaye_PRA2018}.  

Finally, we have shown that our numerical model may be used to simulate the global characteristics of the two-qubit quantum gates such as the truth table and process matrix. We have performed a simulation of full quantum process tomography of a two-qubit gate by calculating the output density matrices for varying input states. The obtained results show that fidelities at the level of 99\% are in principle achievable without any significant modifications to the original blockade gate protocol and this bound may be further shifted by technical improvements, such as utilizing an intermediate state with longer lifetime and aiming for Rydberg states with higher principal quantum number to reduce blockade leakage errors. 


\acknowledgements

\noindent This work was supported by the Russian Science Foundation under
Grant No. 18-72-10039. R.R.Y. acknowledges support from the Foundation for Assistance to Small Innovative Enterprises under Grant UMNIK. S.P.K and S.S.S. acknowledge support by the Interdisciplinary Scientific and Educational School of Moscow University Photonic and Quantum Technologies. Digital Medicine. D.V.K. and C.I.S. acknowledge support by the National Science Foundation under Grant No. 1606743. This work was supported by the Russian Roadmap on Quantum Computing.

\appendix

\section{Focused Gaussian mode}\label{Appendix_A}
\noindent Any of the light beams illuminating the atoms in the paraxial approximation can be described by an appropriate superposition of the mode functions $U^{(s)}(\mathbf{r})$ where the combined mode index ``$s=k;p,l$'' corresponds to the standard Laugerre-Gaussian parameterization with an azimuthal number $l$ and with a radial index $p\geq l$. We consider the case of a fundamental Gaussian mode with $p=l=0$ (TEM${}_{00}$-mode) such that $\omega\equiv\omega_{s}=\omega_{k}=ck$, and the mode function $U^{(s)}(\mathbf{r})$ can be expressed in cylindrical coordinates with the origin at the focal point as
\begin{equation}
U^{(s)}(\mathbf{r})=\frac{1}{\sqrt{{\cal L}}}\,\mathrm{e}^{ikz}\,u(\rho,z)=\mathrm{const}_{\phi},%
\label{a.1}
\end{equation}
where ${\cal L}$ is the quantization length for the periodic boundary conditions, and the slowly varying amplitude is given by
\begin{equation}
u(\rho,z)=a(z)\exp\left[i\frac{k}{2q(z)}\rho^2+i\psi(z)\right]%
\label{a.2}%
\end{equation}
with
\begin{equation}
\frac{1}{q(z)}=\frac{1}{R(z)}+i\frac{\lambda}{\pi\,\mathrm{w}^2(z)},%
\label{a.3}%
\end{equation}
where $R=R(z)$ is the wavefront curvature
\begin{equation}
R(z)=z\left[1+\left(\frac{\pi\,\mathrm{w}_0^2}{\lambda\,z}\right)^2\right],%
\label{a.4}%
\end{equation}
and 
\begin{eqnarray}
\mathrm{w}(z)&=&\mathrm{w}_0\left[1+\left(\frac{\lambda\,z}{\pi\,\mathrm{w}_0^2}\right)^2\right]^{1/2}
\nonumber\\%
\psi(z)&=&\arctan\left(\frac{\lambda\,z}{\pi\,\mathrm{w}_0^2}\right)%
\label{a.5}%
\end{eqnarray}
are the beam waist and the phase shift, respectively (both dependent on $z$). The outer factor $a(z)$ is the normalization constant. The extra phase $\psi(z)$, varying from $-\pi/2$ to $\pi/2$, is known as the Gouy phase and reveals a phase inversion at the beam edges. The important longitudinal scale $z_{\mathrm{R}}=\pi\,\mathrm{w}_0^2/\lambda$, called the Rayleigh range, indicates a length of the beam divergence near the caustic waist. $R(z)$ denotes the caustic curvature in the $\rho,z$ plane and it approaches infinity at $z\to 0$, where we have
\begin{equation}
u(\rho)\equiv u(\rho,0)=\sqrt{\frac{2}{\pi\,\mathrm{w}_0^2}}%
\exp\left[-\frac{\rho^2}{\mathrm{w}_0^2}\right].%
\label{a.6}
\end{equation}
The beam has diffraction limited divergence inside a cone with the polar angle $\theta=\lambda/(\pi\,\mathrm{w}_0)$ and the solid angle associated with the mode is given by $\pi\theta^2=\lambda^2/\pi\,\mathrm{w}_0^2$.

In the paper, we use the following expansion for the profile of the field amplitude near the frame origin coinciding with the caustic focal point
\begin{equation}
 u(\rho,z)/u(0,0) \approx 1-\frac{\rho^2}{w_0^2}-\frac{z^2}{2z_R^2}+i\frac{z}{z_R}+\ldots%
 \label{a.7}%
\end{equation}
which parameterizes the spatial dependence of the Rabi frequencies for both Rydberg excitation beams. \\

\section{Two-photon excitation by plane waves}\label{Appendix_B}

\noindent Once we neglect the differential terms in equations (\ref{2.8}) and approximate the two-photon excitation by spatially homogeneous plane waves, as assumed in (\ref{2.11}), the general solution (\ref{2.12}) simplifies as follows
\begin{equation}
\left(\begin{array}{c}c_{r\mathbf{p}+\hbar\mathbf{q}}(\tau) \\ c_{b\mathbf{p}}(\tau) \\ c_{a\mathbf{p}}(\tau) \end{array} \right)
=\hat{{\cal U}}(\tau)
\left(\begin{array}{c}c_{r\mathbf{p}+\hbar\mathbf{q}}(0) \\ c_{b\mathbf{p}}(0) \\ c_{a\mathbf{p}}(0) \end{array} \right),
\label{b.1}%
\end{equation}
where $\hat{{\cal U}}(\tau)$ is now expressed by a $3\times 3$ matrix, with c-number matrix elements. Straightforwardly we obtain
\begin{widetext}
\begin{equation}
\hat{{\cal U}}(\tau)=\left[\begin{array}{ccc}\left[\cos\left(\displaystyle\frac{\Omega_{\mathbf{p}} \tau}{2}\right)%
+i\displaystyle\frac{\Delta_{\mathbf{p}}}{\Omega_{\mathbf{p}}}\sin\left(\displaystyle\frac{\Omega_{\mathbf{v}} \tau}{2}\right)\right]\,%
\mathrm{e}^{-i\Delta_{\mathbf{p}}\tau/2}%
&i\displaystyle\frac{|\Omega|}{\Omega_{\mathbf{p}}}\sin\left(\displaystyle\frac{\Omega_{\mathbf{p}} \tau}{2}\right)\,\mathrm{e}^{+i\phi-i\Delta_{\mathbf{p}}\tau/2}&0\\ \\%
i\displaystyle\frac{|\Omega|}{\Omega_{\mathbf{p}}}\sin\left(\displaystyle\frac{\Omega_{\mathbf{p}} \tau}{2}\right)\,\mathrm{e}^{-i\phi+i\Delta_{\mathbf{p}}\tau/2}%
&\left[\cos\left(\displaystyle\frac{\Omega_{\mathbf{p}} \tau}{2}\right)-i\displaystyle\frac{\Delta_{\mathbf{p}}}{\Omega_{\mathbf{p}}}\sin\left(\displaystyle\frac{\Omega_{\mathbf{p}}\tau}{2}\right)\right]\,%
\mathrm{e}^{i\Delta_{\mathbf{p}}\tau/2}&0\\ \\
0&0& 1 \end{array}\right]%
\label{b.2}%
\end{equation}
\end{widetext}
where 
\begin{equation}
\Delta_{\mathbf{p}}=\omega_1+\omega_2 -\tilde{\omega}_{rb}-\frac{\mathbf{q}\!\cdot\!\mathbf{p}}{m}-\frac{\hbar\mathbf{q}^2}{2m}%
\label{b.3}%
\end{equation}
or
\begin{equation}
\Delta_{\mathbf{p}}=\omega_1+\omega_2 -\tilde{\omega}_{rb}-\frac{\mathbf{q}\!\cdot\!\mathbf{p}}{m}-\frac{\hbar\mathbf{q}^2}{2m}-\delta_R%
\label{b.4}%
\end{equation}
in the case of blocked excitation i.e. if the control atom $A$ is already in the Rydberg state and the transformation (\ref{b.2}) affects only the target atom $B$.

Here we have denoted $\Omega=|\Omega|\mathrm{e}^{i\phi}$ and defined the generalized Rabi frequency
\begin{equation}
\Omega_{\mathbf{p}}=\sqrt{|\Omega|^2+\Delta_{\mathbf{p}}^2}.%
\label{b.5}%
\end{equation}
As follows from (\ref{b.2}) and (\ref{b.4}) the transition amplitude for the simultaneous excitation of two atoms is suppressed by a factor $|\Omega|/\delta_R$. 

\section{Diagram images of the scattering channels}\label{Appendix_C}

\noindent Identification and classification of the interaction channels with the environment can be relevantly done by the non-equilibrium diagram method introduced by L.M.~Keldysh, see \cite{Keldysh,LfPtX,KUPRIYANOV20171}. To clarify this let us consider the following two-particle correlation function
\begin{eqnarray}
\lefteqn{i{\cal G}_{\alpha'\beta';\alpha,\beta}^{(--++)}(\mathbf{r}'_A,t'_A,\mathbf{r}'_B,t'_B;\mathbf{r}_A,t_A,\mathbf{r}_B,t_B)}
\nonumber\\%
&&= \left\langle\tilde{T}\left[\hat{\Psi}_{\alpha}^{\dagger}(\mathbf{r}_A,t_A)\,\hat{\Psi}_{\beta}^{\dagger}(\mathbf{r}_B,t_B)\right]\right.%
\nonumber\\%
&&\phantom{i{\cal G}_{\alpha'\beta';\alpha,\beta}^{(--++)}}\left.\times T\left[\hat{\Psi}_{\beta'}(\mathbf{r}'_B,t'_B)\,\hat{\Psi}_{\alpha'}(\mathbf{r}'_A,t'_A)\right]\right\rangle%
\label{c.1}%
\end{eqnarray}
where we have used the second quantized formalism, and the chronological operators $T$ and $\tilde{T}$ respectively order and anti-order the product of the system operators in the square brackets in time. The atoms are assumed to be immobile and distinguishable particles. $\hat{\Psi}_{\alpha}^{\dagger}(\mathbf{r}_A,t_A),\,\hat{\Psi}_{\alpha'}(\mathbf{r}'_A,t'_A)\ldots$ are respectively the creation and annihilation operators for atoms $A$ and $B$ at certain spatial points and times evolving in the Heisenberg picture.   

Once we fix the position of each atom and neglect its uncertainty within the trap scale, we can link this correlation function to the two-particle density matrix introduced in the main text 
\begin{eqnarray}
\lefteqn{i{\cal G}_{\alpha'\beta';\alpha,\beta}^{(--++)}(\mathbf{r}'_A,t,\mathbf{r}'_B,t;\mathbf{r}_A,t,\mathbf{r}_B,t)}
\nonumber\\
&&=\rho_{\alpha',\beta';\alpha,\beta}(t)\,\delta(\mathbf{r}'_A-\mathbf{r}_A)\,\delta(\mathbf{r}'_B-\mathbf{r}_B)
\label{c.2}
\end{eqnarray}
Nevertheless it is more convenient to manipulate with (\ref{c.1}), which can be expanded by the diagram series in accordance with conventional rules of the invariant perturbation theory. To follow this concept we can transform (\ref{c.1}) to the interaction picture 
\begin{eqnarray}
(\ref{c.1})&=&\left\langle\tilde{T}\left[\hat{S}^{\dagger}\hat{\Psi}_{\alpha}^{(0)\dagger}(\mathbf{r}_A,t_A)\,\hat{\Psi}_{\beta}^{(0)\dagger}(\mathbf{r}_B,t_B\right]\right.%
\nonumber\\%
&&\times\left.T\left[\hat{\Psi}_{\beta'}^{(0)}(\mathbf{r}'_B,t'_B)\,\hat{\Psi}_{\alpha'}^{(0)}(\mathbf{r}'_A,t'_A)\,\hat{S}\right]\right\rangle%
\label{c.3}
\end{eqnarray}
where $\hat{S}$ denotes the evolution operator
\begin{equation}
\hat{S}=T\exp{\left[-\frac{i}{\hbar}\int_{-\infty}^{\infty}\hat{V}^{(0)}(t)\,dt\right]}.%
\label{c.4}%
\end{equation}
In (\ref{c.3}) and (\ref{c.4}) the operators, superscribed by $(0)$ index, are considered in the interaction picture and we have included the interactions with the external coherent and quantized field modes in the interaction Hamiltonian $\hat{V}^{(0)}(t)$. The expansion of the evolution operators can be regrouped in such a way that it generates multiple partial contributions which can be mapped onto specific diagram images. 

For further details of the Keldysh's diagram approach we refer to the papers cited above. The crucial feature is that the interaction terms generated by the expansion of $\hat{S}$ and contributing to the vacuum expectation values are marked by a $-$ sign but the similar terms generated by the expansion of $\hat{S}^{\dagger}$ -- by a $+$ sign. Below we present the diagram images of the processes described in section \ref{Section_IIIB}.

The depopulation terms in the evolution of the density matrix in Eqs.~(\ref{3.9}) -- (\ref{3.12}) for the light scattering from the states $|a\rangle$ and $|b\rangle$ can be recovered by decoding the following diagrams    
\begin{equation}
\scalebox{0.55}{\includegraphics*{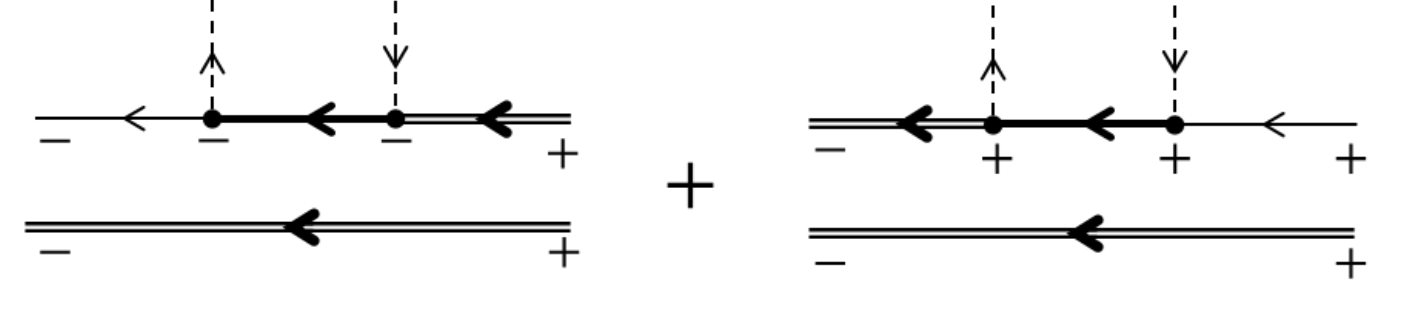}}
\label{c.5}%
\end{equation}
where double lines visualize the original Green's functions of the atoms, undisturbed by incoherent losses but subject to the coherent dynamics, and the thick solid line is the atomic propagator in the intermediate state, dressed by interaction with the vacuum modes. The thin lines are the free propagators in the final state. The dashed arrows here express the interactions with the coherent mode $\omega_1$. Similarly for the scattering from the Rydberg level $|r\rangle$ we obtain 
\begin{equation}
\scalebox{0.55}{\includegraphics*{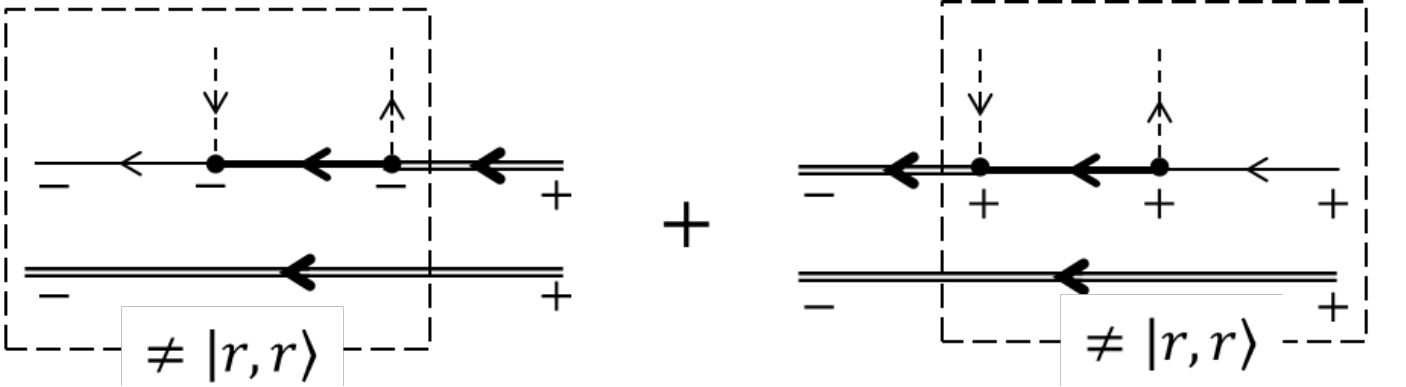}}
\label{c.6}%
\end{equation}
where the dashed arrows express the interactions with the coherent mode $\omega_2$. In evaluation of these diagrams, as well as of (\ref{c.9}) and (\ref{c.10}) below, we extract only the spontaneous contributions, proportional to $\gamma$, and treat them as a small perturbation. The dynamical behavior of the process is already incorporated into the double arrow lines.\footnote{\noindent The atomic propagator in the intermediate states, represented by the thick solid line, is $\hbar(E-E_n\pm i\gamma/2)^{-1}$, where the sign depends on type of the time ordering. These diagrams reproduce the effective Hamiltonian discussed in section \ref{Section_IIA} if $\gamma$ is neglected. The first order corrections with respect to $\gamma$ then describe the incoherent depopulation processes.} Note that we have dealt here with an entangled pair of atoms and have excluded those events when both atoms originally occupy the Rydberg state. The probability of such an event is small and beyond the approximations made. Here and below we point out the latter circumstance by dashed-boxing the forbidden processes in the diagrams.  To obtain other depopulation diagrams visualizing the damping of coherency, originally created by dynamical interaction between the ground and Rydberg states, one has to combine the cross parts from (\ref{c.5}) and (\ref{c.6}). 

The repopulation of atoms by optical pumping, see Eq.~(\ref{3.13}), is expressed by the diagrams
\begin{equation}
\scalebox{0.4}{\includegraphics*{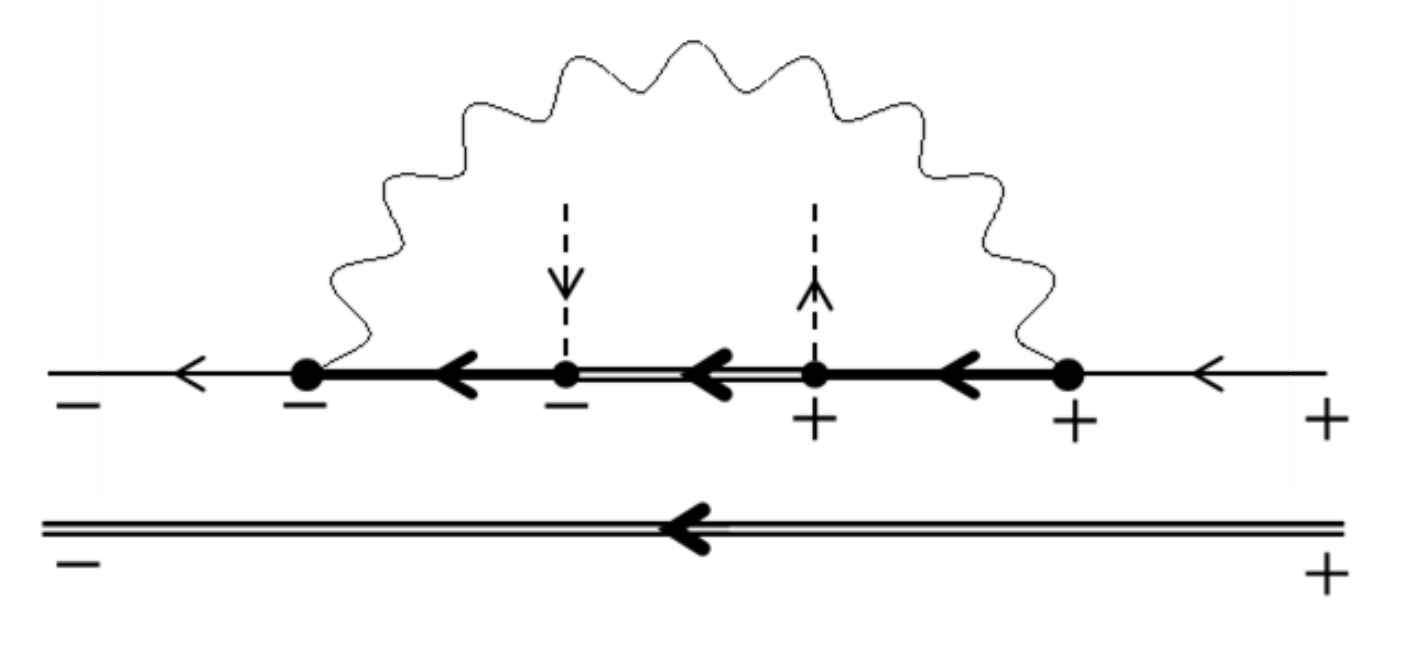}}
\label{c.7}%
\end{equation}
for the scattering from the ground state and 
\begin{equation}
\scalebox{0.4}{\includegraphics*{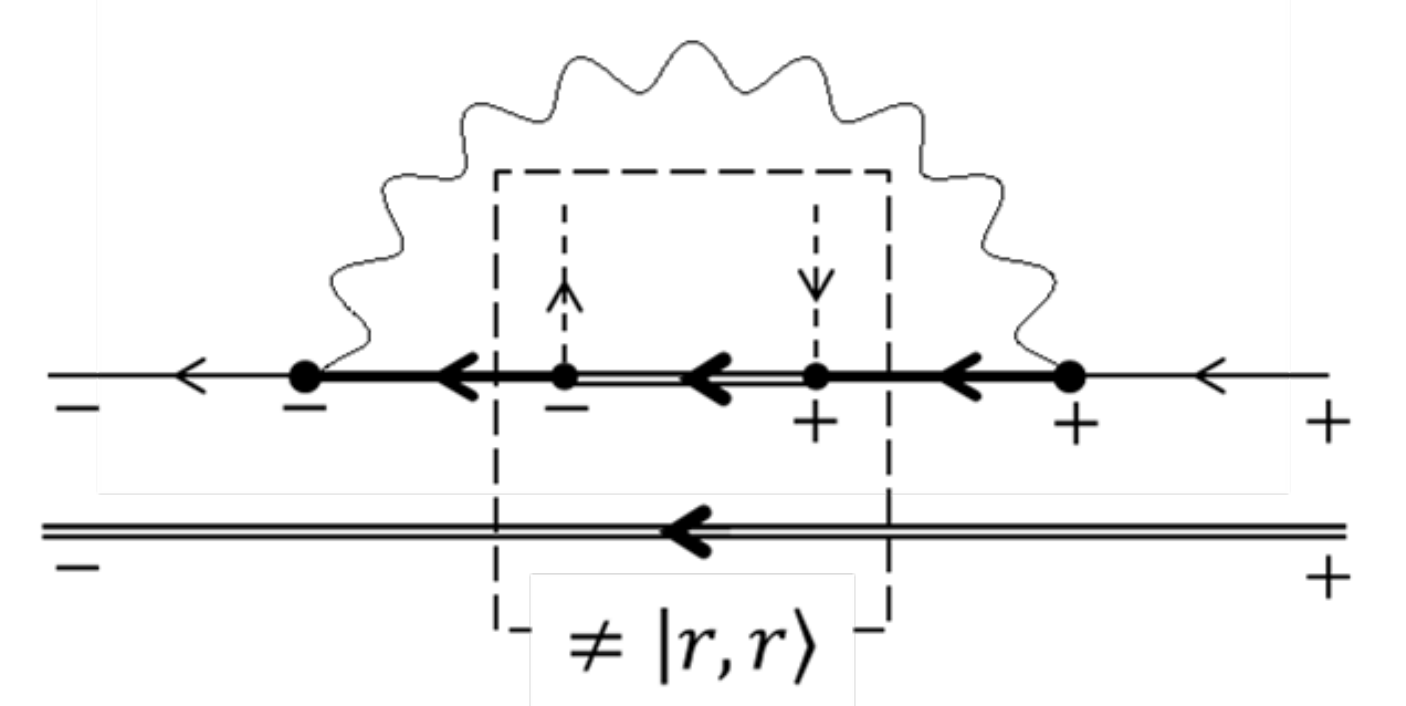}}
\label{c.8}%
\end{equation}
for the scattering from the Rydberg state. Here the photon's wavy line indicates tracing over all scattering directions of the emitted photon.

The depopulation processes, induced by the two-photon resonance, see Eqs.~(\ref{3.14}), (\ref{3.15}) and (\ref{3.17}) are imaged by the following diagrams
\begin{equation}
\scalebox{0.55}{\includegraphics*{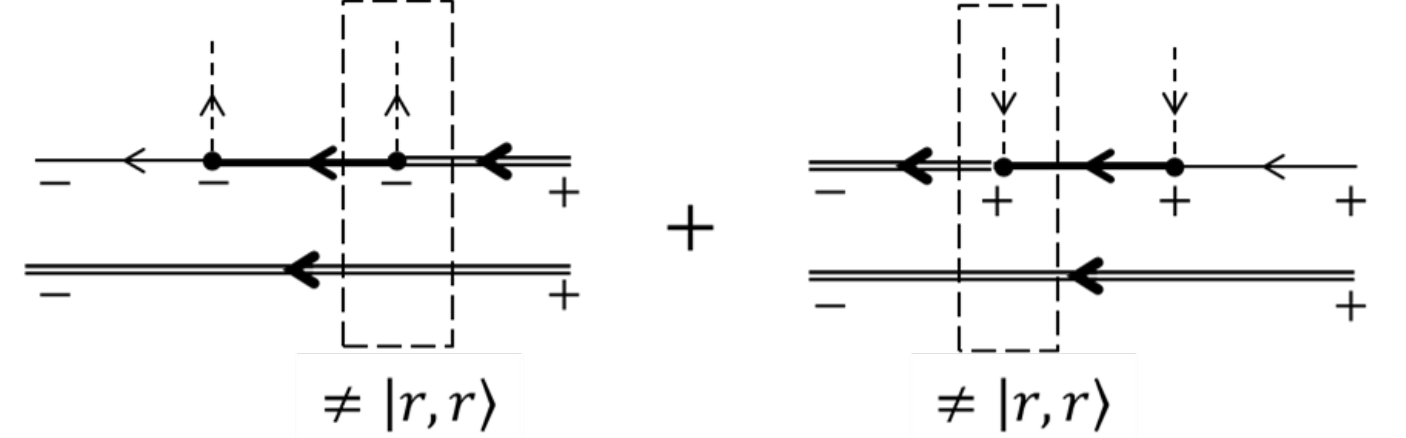}}
\label{c.9}%
\end{equation}
and
\begin{equation}
\scalebox{0.55}{\includegraphics*{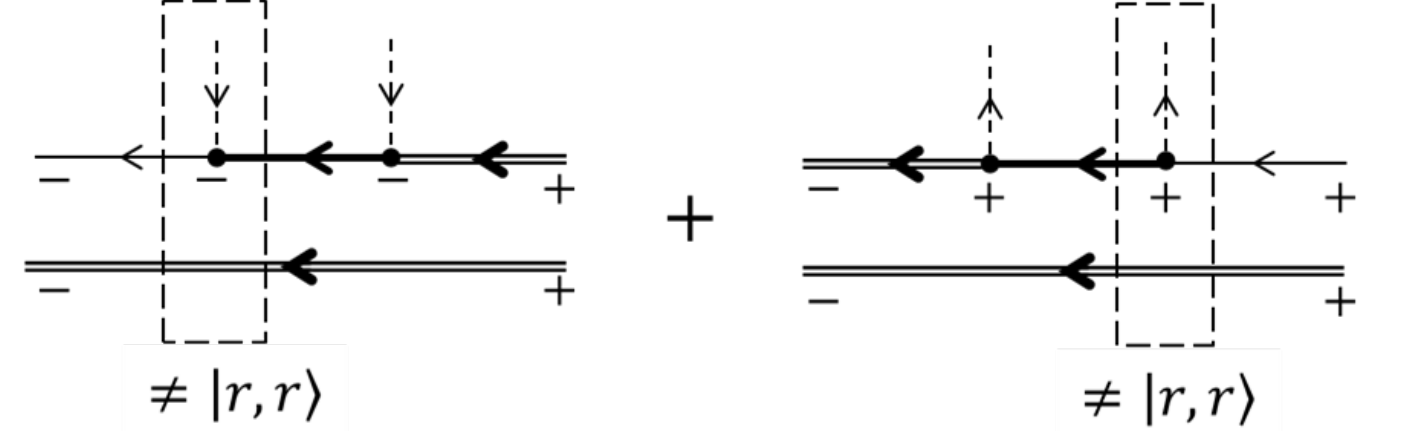}}
\label{c.10}%
\end{equation}
and one has to combine the cross parts from (\ref{c.9}) and (\ref{c.10}) to obtain the diagram visualizing the damping of Rydberg coherence.

The specific repopulation terms, induced by the Rydberg coherences (\ref{3.16}) and (\ref{3.18}), are imaged by the diagrams
\begin{equation}
\scalebox{0.4}{\includegraphics*{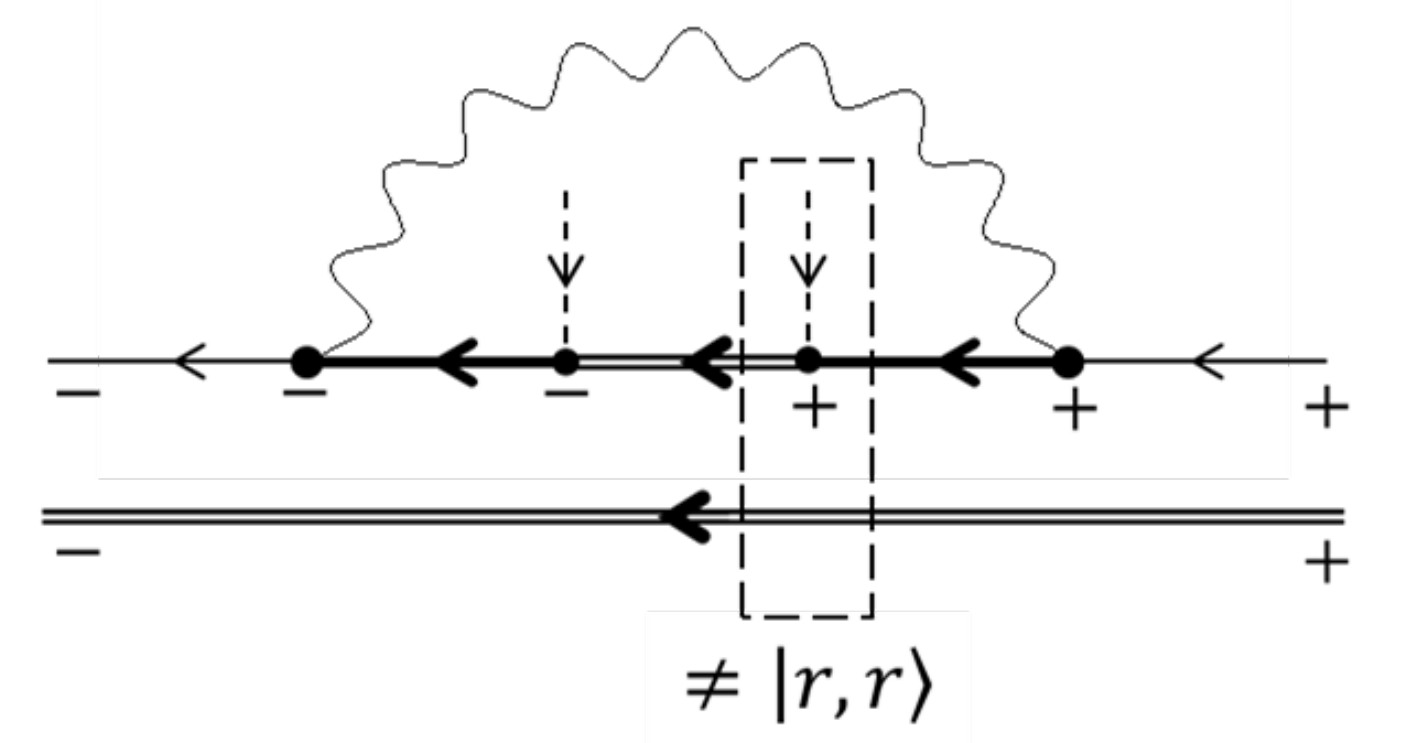}}
\label{c.11}%
\end{equation}
and
\begin{equation}
\scalebox{0.4}{\includegraphics*{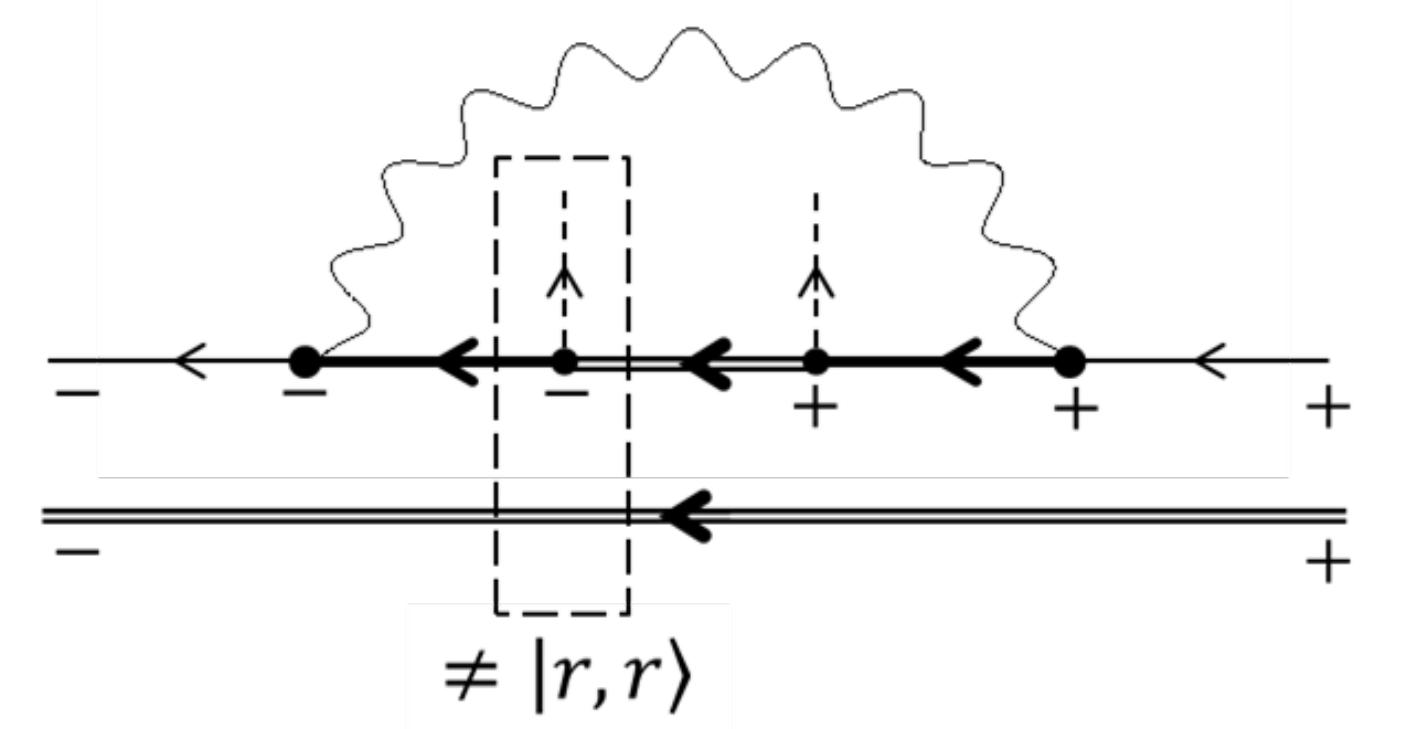}}
\label{c.12}%
\end{equation}
where (\ref{3.16}) is given by the decoded sum of both the graphs, but (\ref{3.18}) is given by (\ref{c.12}) when decoding of (\ref{c.11}) gives its Hermitian conjugated counterpart.

\section{Reconstruction of the $\chi$-matrix for the CNOT gate}\label{Appendix_D}

\noindent When a realistic simulation of the entangling gate is obtained, it can be used to simulate the procedure of the quantum process tomography aiming at providing the most detailed description of the underlying quantum process. In the most general case an arbitrary quantum transformation $\mathcal{E}(\cdot)$ can be rigorously described by making use of the so called $\chi$-matrix or process matrix defined as follows
\begin{equation}
\mathcal{E}(\rho^{\mathrm{in}}) = \sum_{m, n}^{D^2} \tilde E_m \rho^{\mathrm{in}}\tilde E_n ^\dagger\, \chi_{m, n}
\label{d.1}%
\end{equation}
where $\rho^{\mathrm{in}}$ stands for the initial or the input state density matrix, $D$ is the dimension of the system's state space. For the sake of notation convenience in this Appendix we specify arbitrary basis states and the linear operators defined in the unitary space, being a linear span of the original computation atomic basis $|\alpha,\beta\rangle$ with $\alpha,\beta=a,b$, by integer numbers and by their compositions, see definitions in the main text. 

$\tilde{E}_n$ and $\tilde{E}_m$ denote a set of basis operators acting in this space, such that (\ref{d.1}) can be rewritten as 
\begin{equation}
\mathcal{E}(\rho^{\mathrm{in}}) = \sum_{i}^{D^2} E_i \rho^{\mathrm{in}} E_i^\dagger, 
\label{d.2}%
\end{equation}
where
\begin{equation}
E_i = \sum_{m}^{D^2}e_{im} \tilde E_m.
\label{d.3}%
\end{equation}
In general the basis operators $\tilde E_m$ can be chosen arbitrarily, but it is convenient for our purposes to define them as the following dyad-type operators 
\begin{equation}
\tilde E_m = E_{ m_1,m_2} = |m_1\rangle\langle m_2|
\label{d.4}%
\end{equation}
where we have implied the composite notation $m=m_1,m_2$, where $m$ can be further enumerated by an integer number running from $1$ to $D^2$. The expression (\ref{d.2}) generates a set of transformation matrices for any evolution process by varying the expansion coefficients $e_{im}$. We accumulate the details of the evolution process in the matrix $\chi$ by transforming from (\ref{d.2}) to (\ref{d.1}).

Let us substitute $\mathcal{E}(\rho^{\mathrm{in}}) = \rho^{\mathrm{out}}$ and then select an arbitrary matrix element $\rho^{\mathrm{out}}_{j, k}$, taken in the original basis. Then we arrive at 
\begin{eqnarray}
\rho^{\mathrm{out}}_{j, k} &= &\sum_{m_1, m_2, n_1, n_2}^D \langle j|m_1\rangle\langle m_2|\rho^{\mathrm{in}}|n_2\rangle\langle n_1|k\rangle \tilde\chi_{m_1, m_2; n_1, n_2}%
\nonumber\\%
&&\hspace{1cm}= \sum_{m_2, n_2}^{D} \rho^{\mathrm{in}}_{m_2, n_2} \tilde\chi_{j, m_2; k, n_2}%
\label{d.5}%
\end{eqnarray}
where the super-matrix $\tilde\chi$, being rearranged in normal square-matrix representation, conventionally transforms to the process matrix. Here $j=m_1$ and $k=n_1$, so we obtain 
$$\tilde\chi_{m_1, m_2; n_1, n_2}\equiv \chi_{m, n}$$ 
and enumerate the matrix elements as
$$m = D \cdot m_1 + m_2;\ \ 
 n = D \cdot n_1 + n_2$$
by definition. There are many ways to convert the super-matrix $\tilde\chi$ to a square-matrix, so we use the original state specification for the $\chi$-matrix formalism to avoid any uncertainty. 

\begin{figure}[tp]
\includegraphics[width=8.2cm]{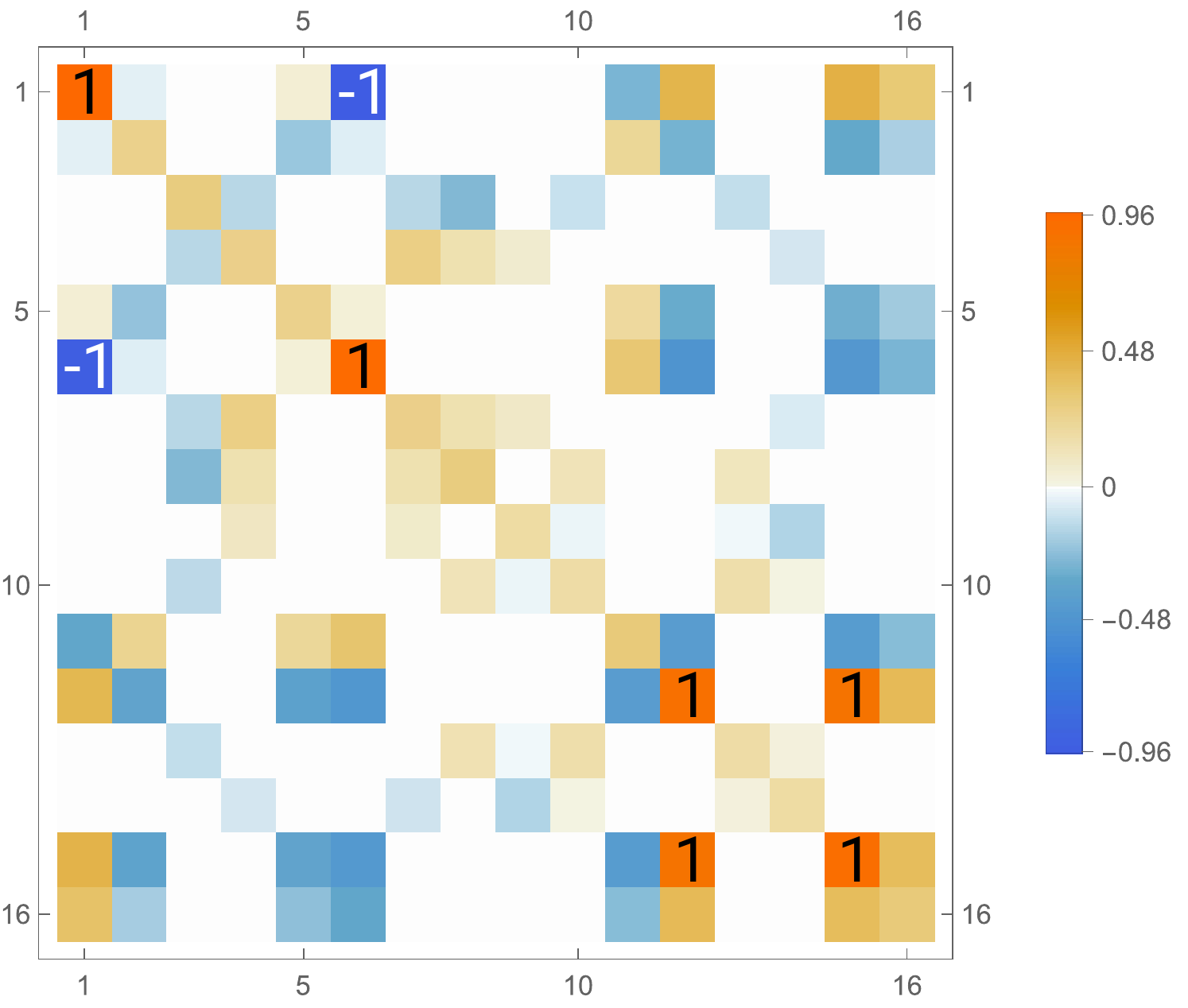}
\caption{Real part of the $\chi$-matrix recovered for the simulated CNOT process, see the main text. }
\label{fig10}
\end{figure}

\begin{figure}[tp]
\includegraphics[width=8.2cm]{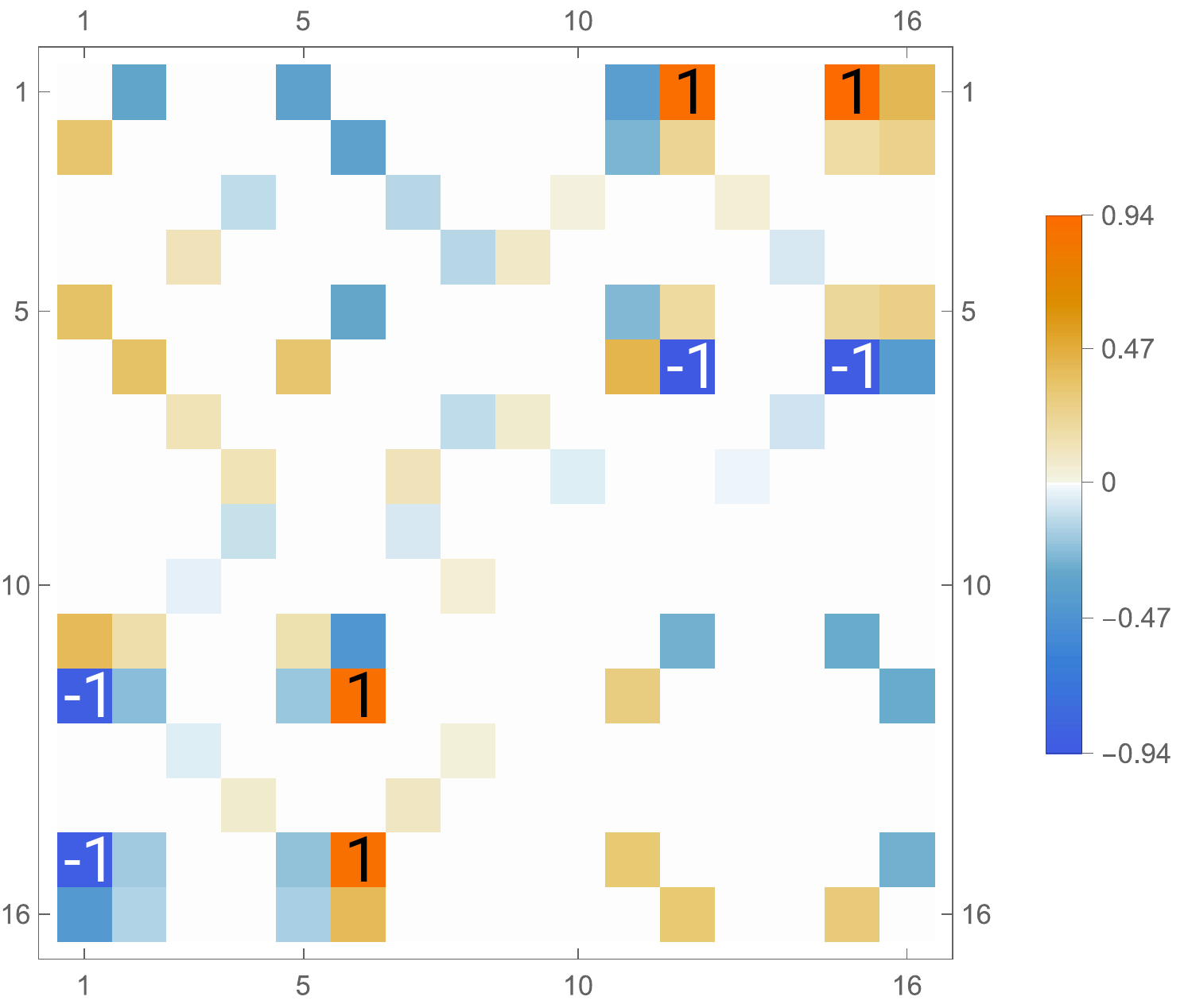}
\caption{Imaginary part of the $\chi$-matrix recovered for the simulated CNOT process, see the main text. }
\label{fig11}
\end{figure}

Equation \ref{d.5} can be resolved and the process matrix can be recovered for any physical realization of the CNOT protocol. That allows us to implement universal process tomography and eligible verification of the gate realization. In order to show this the CNOT simulation was repeated for sixteen different input states from the state space of two coupled qubits. The input states were constructed from the combinations of $|a\rangle$, $|b\rangle$, $(|a\rangle + |b\rangle)/\sqrt{2}$ and $(|a\rangle + i|b\rangle)/\sqrt{2}$ for both the control and the target qubits. Finally, the process super-matrix $\tilde{\chi}$ was recovered as a solution of equation (\ref{d.5}) considered for sixteen different realizations of the pure input and mixed output states. 

\begin{figure}[tp]
\includegraphics[width=8.6cm]{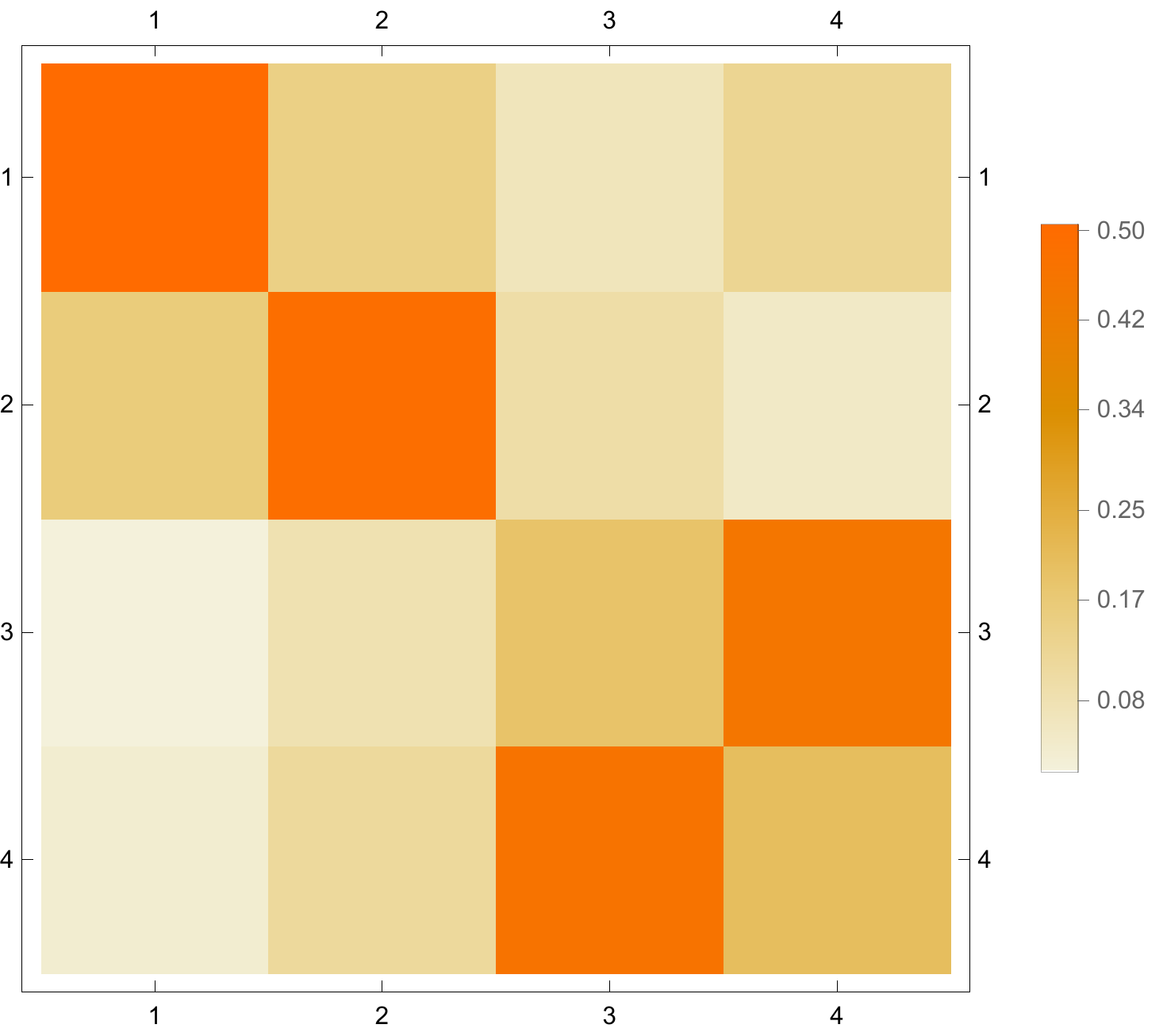}
\caption{Absolute values of the eigenvector corresponding to the largest eigenvalue of the CNOT $\chi$-matrix rearranged as a $4\times4$ matrix.}
\label{fig12}
\end{figure}

As an illustrative example, the process $\chi$-matrix was recovered for the excitation geometry of Fig. \ref{fig3} and the results are shown in Figs.~\ref{fig10} and \ref{fig11}. In these tables we have highlighted the elements of the $\chi$-matrix which have non-zero values for the case of an ideal CNOT protocol by number indicators. Uncolored empty cells contain zero matrix elements.  The process matrix is Hermitian and positively defined by construction. Its largest eigenvector (having a maximal eigenvalue), being rearranged as a four-by-four matrix, is the unitary transformation which is the closest one to the reconstructed process. It is shown in Fig.~\ref{fig12} and is equivalent to an alternative estimate of the truth table shown in Figs.~\ref{fig7},\ref{fig9} in the main text. It reasonably resembles the expected physical realization of the unitary CNOT gate.

\bibliography{references}

\end{document}